\documentclass{aa}  
\usepackage{graphicx}
\usepackage{amsmath}
\usepackage{natbib}
\usepackage{xcolor}
\usepackage{float}
\usepackage{caption}
\usepackage{subcaption}
\usepackage{multirow}
\usepackage[section]{placeins}
\usepackage[dvipsnames]{xcolor}
\usepackage[linkcolor=black,citecolor=blue,filecolor=black,urlcolor=magenta]{hyperref}
\usepackage{txfonts}
\usepackage{hyperref}
\hypersetup{
    colorlinks=true,
    linkcolor=blue,
    citecolor=Cerulean,
    filecolor=magenta,      
    urlcolor=magenta,
    pdftitle={Overleaf Example},
    pdfpagemode=FullScreen,
    }
    
\usepackage{xspace}

\newcommand{\cii}{[\textrm{CII}]\xspace}
\newcommand{\alphacii}{\ensuremath{\alpha_{\textrm{\cii}}}\xspace}
\newcommand{\dgr}{\ensuremath{\rm{\delta_{GD}}\xspace}}
\newcommand{\alphaco}{\ensuremath{\alpha_{\textrm{CO}}}\xspace}
\newcommand{\Lcii}{\ensuremath{L_{\cii}}\xspace}     
\newcommand{\Lsun}{\ensuremath{L_\odot}\xspace}
\newcommand{\Mstar}{\ensuremath{M_\ast}\xspace}     
\newcommand{\Msun}{\ensuremath{M_\odot}\xspace}
\newcommand{\fg}{\ensuremath{f_{mol}}\xspace}

\newcommand{\LIR}{\ensuremath{\rm{LIR_{tot}}}\xspace}
\newcommand{\zspec}{\ensuremath{z_{\textrm{spec}}}}

\newcommand{\Ha}{\textrm{H}\ensuremath{\alpha}\xspace}
\newcommand{\Hb}{\textrm{H}\ensuremath{\beta}\xspace}

\newcommand{\OIII}{[\textrm{O}~\textsc{iii}]\xspace}

\begin{document} 

\title{A first [CII] view of high-z quiescent galaxies}

\author{C.~D'Eugenio \inst{1,2} 
\and E.~Daddi\inst{2}
\and R.~Gobat\inst{3}
\and S.~Jin\inst{4,5} 
\and D.~Liu\inst{6,7}
\and H.~Sun\inst{8,9}
\and F.~Gentile\inst{2}
\and F.~Bruckmann\inst{2}
\and Z.~Liu\inst{2,10,11,12}
\and I.~Delvecchio\inst{13}
\and L.~Vallini\inst{13}
\and B.~Magnelli\inst{2}
\and A.~Zanella\inst{13}
}

\institute{Institute de Physique du Globe de Paris, 5, Rue Jussieu, Paris, France
\and CEA, IRFU, DAp, AIM, Université Paris-Saclay, Université Paris Cité, Sorbonne Paris Cité, CNRS, 91191 Gif-sur-Yvette, France
\and Instituto de Física, Pontificia Universidad Católica de Valparaíso, Casilla 4059, Valparaíso, Chile
\and Cosmic Dawn Center (DAWN), Denmark
\and DTU-Space, Technical University of Denmark, Elektrovej 327, DK2800 Kgs. Lyngby, Denmark
\and Max-Planck-Institut für extraterrestrische Physik (MPE), Giessenbachstrasse 1, D-85748 Garching, Germany
\and Purple Mountain Observatory, Chinese Academy of Sciences, 10 Yuanhua Road, Nanjing 210023, China
\and School of Astronomy and Space Science, Nanjing University, Nanjing 210093, China
\and Key Laboratory of Modern Astronomy and Astrophysics, Nanjing University, Ministry of Education, Nanjing 210093, China
\and Kavli Institute for the Physics and Mathematics of the Universe (Kavli IPMU, WPI), UTIAS, The University of Tokyo, Kashiwa, Chiba 277-8583, Japan
\and Department of Astronomy, School of Science, The University of Tokyo, 7-3-1 Hongo, Bunkyo, Tokyo 113-0033, Japan
\and Center for Data-Driven Discovery, Kavli IPMU (WPI), UTIAS, The University of Tokyo, Kashiwa, Chiba 277-8583, Japan
\and INAF – Osservatorio di Astrofisica e Scienza dello Spazio di Bologna, Via Gobetti 93/3, I-40129 Bologna, Italy
}

\offprints{C. D'Eugenio, \email{cdeugenio1@gmail.com}}

\date{Received XX YY, 2026; accepted XXX YYY, 2026}

\abstract
%   % {} leave it empty if necessary  
{%Recent observations indicate that star formation quenching in massive galaxies starts already a few hundred million years after the Big Bang. Constraining the underlying quenching mechanisms involved requires characterizing the residual interstellar medium \textbf{(ISM)} in high-z quenched galaxies which, however, has long been \textbf{mostly} out of reach due to the prohibitive integration times required in millimeter bands. Here 
We present ALMA detections (or stringent upper limits) of the [CII] 158 $\mu m$ emission line and underlying dust continuum from five massive quenched galaxies (QGs) at 2<z<4.7.
%in synergy with JWST/NIRCam, MIRI and VLA imaging.
%Under the hypothesis that  conversion factor applies to our sample, 
We find extreme variations in the molecular gas fractions (\fg=$M_{mol}$/\Mstar), spanning 0.1\%-25\%, if a standard \alphacii applies. We attempt a first empirical calibration of \alphacii with 
respect to dust continuum in a $z=2$ lensed QG and 
%using one of the most distant dust continuum–detected QG at Cosmic Noon and the only QG 
with respect to CO(3-2) in a $z=3.1$ QG, finding no evidence of strong deviations from the standard value. Dust continuum measurements, coupled with JWST/MIRI fluxes, suggest higher dust temperatures compared to expectations from $z<2$ QGs, reaching $T_{d}\sim40-50 \,K$ in two galaxies. %exceeding expectations from simple \textbf{size evolution models where cirrus dust clouds are exposed to a stronger interstellar radiation field as galaxies become more compact with redshift.} 
Coupled with remarkably high total infrared luminosities (LIR) not explained by observed JWST colors not by energy balance based on literature dust extinction measurements, and with [CII] deficits down to $\rm{[CII]/LIR\sim 2\times10^{-4}}$ typical of (Ultra)Luminous Infrared Galaxies, our findings point to additional dust-heating mechanisms other than dust-absorbed stellar radiation.
%These QGs  display  
Surprisingly, JWST/NIRCam and ALMA imaging reveal widespread disturbed stellar morphologies and offsets/tails in dust and gas, indicative of ongoing interactions.  
While larger samples are needed to assess how common these features are in high-z QGs, these findings support a merger-driven origin for the phenomenology observed in these systems, with key similarities with respect to local post-starburst galaxies where low-velocity shocks and turbulence also inject energy into the residual ISM.
}



\maketitle
\nolinenumbers

\section{Introduction}\label{sec:intro}

In recent years, a wealth of observational evidence has revealed that massive (log(\Mstar/\Msun)>10.5) quenched galaxies (MQGs) emerged remarkably early in cosmic history \citep{Schreiber18b, Valentino20, Carnall23, Tanaka24, deGraaff25, Weibel25, Antwi-Danso25}. The onset of star formation quenching is fundamentally linked to profound changes in the availability or properties of a galaxy's interstellar medium (ISM). This can occur through multiple, not necessarily independent, pathways: gas heating \citep{Ji2024}, rapid consumption \citep{Schawinski2014}, gas loss through outflows \citep{Belli24, Valentino25},  galaxy-galaxy interactions \citep{Khoram25, DOnofrio25}, ram-pressure stripping \citep{Boselli2022, Xu25, Rhee25}, stabilization against gravitational collapse \citep{Martig09, michalowski24} or starvation \citep{Dekel09, Feldmann15}. Probing the residual cold gas content of these galaxies, however, has proven to be challenging. 
Perhaps the greatest observational barrier that the community faces at the time of writing, is the rapid increase in observational time request to probe low gas masses in MQGs at $\rm{z>1}$. Targeting standard $\rm{H_{2}}$ tracers, such as CO(1-0)/(2-1)/(3-2), [CI] and dust continuum, requires strongly lensed galaxies \citep{Whitaker21b}, stacking large galaxy samples \citep{Gobat18, blanquez23, Adscheid2025}, or tens of hours of integration with the most powerful (sub)mm facilities like ALMA and NOEMA \citep{Scholtz24, Umehata2025}.
Additionally, some simulations predict gas-to-dust ratios ($\rm{\delta_{GD}}$) several orders of magnitude larger than in star-forming galaxies \citep{Whitaker21a}, that would make direct observations of dust continuum in MQGs virtually impossible at $z>2$. Recent observations of quenched galaxies at $z\sim1$ appear to support this scenario \citep{Spilker25}.\\
In \cite{deugenio23} we proposed the use of the singly ionized carbon [CII] emission line at 158 $\mu$m as a probe of the cold atomic and molecular gas component of MQGs in the distant Universe. Based on local QGs, we argued that the standard $\rm{\alpha_{[CII]} = 31 \,M_{\odot}/L_{\odot}}$ conversion factor from \cite[][hereafter Z18]{Zanella18}, empirically calibrated on star-forming galaxies at 0 < z < 6 over 4 orders of magnitudes in [CII] luminosities (\Lcii) and molecular gas masses ($\rm{M_{mol}}$), would also likely apply at high redshift. 
This method is best suited for recently quenched or transitioning systems, with sSFRs less than 10\% that of the coeval Main Sequence\footnote{$\rm{SSFR=SFR/M_{\star}\lesssim10^{-10} yr^{-1}}$}, but with enough residual star formation to have enough far-UV (6-13.6 eV) photons produced by OB stars creating Photodissociation Regions (PDRs), whose major coolant is the [CII] line\footnote{We will argue later in the paper that other [CII] powering sources might be at play in the quiescent galaxies detected in our sample.}.
In the local Universe, gas-rich early-type galaxies (ETGs) from the ATLAS$^{\rm 3D}$ survey \citep{cappellari11} exhibit a power-law correlation between $L_{\rm CO(1-0)}$ and $L_{\rm [CII]}$ \citep{Lapham2017, Zhao24}, extending down to ${\rm sSFR}\lesssim3\times10^{-11}{\rm yr^{-1}}$ and SFRs of a few $M_\odot{\rm yr^{-1}}$. This empirical relation supports a connection between [CII] emission and diffuse molecular gas even in systems with strongly suppressed star formation. 
If this assumption on \alphacii is correct, [CII] observations at $3<z<6$ would probe molecular gas fraction limits significantly deeper than achievable with dust continuum or low-$J$ CO transitions at fixed integration time \citep{deugenio23}, allowing us to access the properties of the ISM in the early generations of MQGs. 
A critical step, however, is validating the applicability of a standard $\alpha_{\rm [CII]}$ for the high-z quenched galaxy population. 
Thankfully, ALMA high-frequency Bands 9 and 10, allow us to observe [CII] in galaxies at $\rm{1.2<z<1.4}$ and $\rm{1.8<z<2.0}$, enabling cross-calibration of [CII]-based gas masses against existing dust continuum and CO measurements in spectroscopically confirmed MQGs.

In this work we present the first observations targeting [CII] and its underlying continuum dedicated to MQGs at $\rm{2<z<4.7}$, using ALMA bands 7, 8 and 9, including supporting imaging from JWST. We describe our sample in Sect.~\ref{sec:sample} and our data reduction and analysis in Sect.~\ref{sec:analysis}. We provide results on the [CII] line and continuum measurements in Sect.~\ref{sec:cii_results}, while we constrain the dust temperature of our galaxies through FIR SED fitting in Sect.~\ref{sec:SED_analysis}. We discuss our findings and related uncertainties in Sect.~\ref{sec:discussion}. Finally, we summarize our conclusions in Sect.~\ref{sec:summary}.
%Throughout this work 
We assume a $\Lambda$CDM cosmology with H$_{0}=70$ km s$^{-1}$ Mpc$^{-1}$, $\Omega_{M}=0.27$, $\Omega_{\Lambda}=0.73,$ and a \citet{Chabrier03} initial mass function (IMF). Magnitudes are given in the AB photometric system.

\section{Sample description}\label{sec:sample}
The galaxies analyzed in this work were selected as well-studied and spectroscopically confirmed quiescent galaxies with log(\Mstar/\Msun)>10.5 lying at redshifts compatible with the most favorable transmission windows for high frequency observations of the redshifted [CII] line with ALMA. They pertain to two ALMA observing programs: program ID 2023.1.00609.S (P.I. R. Gobat)  aimed at probing the cold gas budget of high-z MQGs through [CII], thus  focused on galaxies at $\rm{z_{spec}>2.8}$; program ID 2024.1.00099.S (P.I. R. Gobat) aimed at calibrating the \alphacii in MQGs at intermediate redshifts by following-up galaxies with publicly available gas mass measurements from detections either in dust continuum or CO transitions.
Despite only a subset of the pointings in these two programs were observed (33\% and 25\%, respectively), we still found invaluable information. We provide an overview of the completed observations in Table~\ref{table:obs}. In what follows we present observed targets.

\begin{table*}
\caption{Overview of the observing programs used for this work. Values given for the pointing targeting GS-9209 refer to the concatenated measurements sets of programs 2023.1.00609.S and 2023.1.01016.S.}          
\label{table:obs}
\centering  
\resizebox{\textwidth}{!}{\begin{tabular}{c c c c c c c c c c}
\hline\hline         

Program & $z_{spec}$ &  ID$_{source}$ &  ALMA band & $\rm{t_{obs}}$ & rms$_{\rm cont}$ & beam & spectral resolution & MRS \\ 
 & & & &  [h] & [mJy/beam] &  [arcsec $\times$~arcsec] & [km/s] & [arcsec] \\  
\hline               \\
2023.1.00609.S + 2023.1.01016.S & 4.66 & GS-9209 & 7 &6.61 & 0.008 & 0.18 $\times$ 0.16 & 36 & 2.9\\\\

2023.1.00609.S & 3.09 & ADF22 QG1/QG2/QG3 & 8  & 0.91 & 0.093 & 0.52 $\times$ 0.44 & 27 & 4.6\\\\
 
\hline             \\
2024.1.00099.S & 1.95 & M0138 & 9 & 0.39 & 0.24 & 0.54 $\times$ 0.43 & 14 & 4.01\\\\

\hline\hline 
\end{tabular}}
\end{table*}

\begin{itemize}
\item{} \textbf{MRG-M0138} (hereafter M0138) is a massive ($\rm{log(\Mstar/\Msun)=11.7\pm0.2}$), lensed z=1.9486 QG with multiple images \citep{Newman18}. 
The latest measurements of its stellar velocity dispersion and de-lensed star formation rate are $\rm{v_{disp}= 398\pm12\,km\,s^{-1}}$ \citep{Newman2025} and $\rm{SFR\sim2\pm1\,\Msun\,yr^{-1}}$\citep{Newman18}. 
We here focus on Image 1 which was found to host a compact dusty core visible in continuum at 1.3 mm \citep[REQUIEM Survey,][]{Whitaker21a}, which we followed-up in [CII]. 
The dust continuum core displays a flux density of $\rm{S_{1.3mm}=0.27\pm0.03\,mJy}$ associated to a gas fraction of $\rm{f_{gas}=0.6\pm0.1 \%}$, after point-source extraction, assuming a dust temperature of $\rm{T_{d}=25\,K}$ and a gas-to-dust ratio of $\rm{\delta_{GD}=100}$ \citep{Whitaker21a}. In \cite{Gobat22} the stellar and gas masses and their uncertainties were recomputed by extracting their flux using a model of the extended NIR light to account for potential diffused dust distributed outside of the bulge. This method resulted in an increased total flux by about 25\% for this galaxy;

\item \textbf{ADF22-QG1} (hereafter QG1) is a MQG at \zspec=3.092$^{+0.008}_{-0.004}$ displaying strong Balmer absorption lines in the stellar continuum \citep[][hereafter K21]{Kubo21}. It resides, together with QG2 and QG3, in the AzTEC14 galaxy group, in the core region of the SSA22 protocluster, covered by the "ALMA Deep Field in SSA22" (ADF22) \citep{Steidel98, Umehata15}. We adopt JWST-based stellar masses from \cite{Umehata2025}, thus $\rm{\Mstar=1.28\times10^{11}\,\Msun}$ for QG1. 
Its size and Sérsic index have been constrained to $r_{e}=1.01\pm0.04$ kpc ($r_{e}\sim0.13$") and n$_{Sérsic}$=2.5$\pm$0.2, respectively \citep[][K21]{Kubo17}, although new constraints on its morphology have been announced (M. Kubo et al. in preparation). As for its stellar velocity dispersion, we adopt here the best-fitting result in K21, i.e.  $\rm{v_{disp}=320\,km\,s^{-1}}$. 
A 1.2 mm dust continuum $3\sigma$ upper limit of 75 $\mu$Jy was reported for QG1 by K21. This was converted into a total infrared luminosity and SFR upper limits of $\rm{L_{IR}(8\text{–}1000\,\mu m)}<0.9\text{–}2.0\times10^{11}\,L_\odot$ and $\rm{SFR}<9\text{–}21 \,M_\odot yr^{-1}$, respectively. The same upper limit has been assigned also to QG2 and QG3.
QG1 does not show any [OIII] line emission and is not detected nor in X-rays nor in the radio domain. Its [OII] emission line is weak and could be consistent with residual low lever star-formation or with LINER emission from weak AGN activity (K21). Recently, QG1 has been detected in CO(3-2) \cite{Umehata2025}, corresponding to a molecular gas fraction of 17\%, making it the most distant CO-detected QG in the literature;

\item \textbf{ADF22-QG2} (hereafter QG2) is a MQG hosting a strong ionized outflow from a type-2 QSO \citep[][hereafter K22]{Kubo22}. Its MOSFIRE and MOIRCS spectra show broad \Hb and \OIII emission lines but no information on stellar absorption lines. The redshift of this galaxy is constrained to be between z=$3.082^{+0.001}_{-0.001}$ and z=$3.091^{+0.001}_{-0.003}$, depending on the slit used for spectral extraction of the [OII] and [OIII] lines and on which line is considered (K22). The large $\rm{log( \OIII / \Hb)\sim 1.00\pm0.05}$ ratio is similar to optically faint AGN in the same redshift range and to young star-forming galaxies with high ionization parameters at z=2-3, as noted in K22. The galaxy is both an X-ray and a radio source. Using its 1.4 GHz luminosity and the upper limit on $\rm{L_{IR}}$, K22 derive an upper limit on the logarithmic ratio of the total infrared and rest 1.4 GHz luminosity of $\rm{q_{IR}}<1.4$.  QG1 and QG2 lie in the passive region of the UVJ diagram (K22);

\item \textbf{ADF22-QG3} (hereafter QG3) was presented in \cite{Kubo16} as a massive [OIII]-emitter at $\rm{z=3.0774\pm 0.0003}$. Its MOIRCS spectrum consists of an emission line partially covered by an OH sky line, with no stellar continuum detected. QG3 is not detected in 1.2 mm dust continuum, similarly to QG1 and QG2. %QG2 and QG3 show very tentative ($\sim2\sigma$) CO(3-2) emission at the frequency expected from \OIII detections \citep{Umehata2025}.

\item \textbf{GS-9209} is a massive ($\rm{\Mstar=4.1\pm0.2\,10^{10}\, \Msun}$) QG at \zspec=4.6582$^{+0.0002}_{-0.0002}$ \citep{Carnall23}. Its U-band rest-frame effective radius is $\rm{r_{e}\sim200\,pc}$ ($\rm{r_{e}\sim0.03}$"), reaching an extreme stellar surface density of $\rm{log(\Sigma_{eff}/\Msun\,kpc^{-2})=11.16\pm0.08}$. The velocity dispersion is $\rm{v_{disp}=247 \pm 16\,km\,s^{-1}}$. Its JWST/NIRSpec spectrum exhibits a broad \Ha emission line, indicative of a type 1 AGN. The dust attenuation measured from the NIRSpec spectrum is very low $\rm{A_{V} = 0.04_{-0.03}^{+0.05} mag}$. The SFR associated to the \Ha narrow component is $\rm{SFR = 1.9 \pm 0.1 \, \Msun\,yr^{-1}}$ \citep{Carnall23}. This target was also observed in [CII] for 4.435 hours
from program 2023.1.01016.S (P.I. Z. Ji), 
%reaching a $\sigma_{rms}=\textbf{0.027}$ mJy/beam, 
with a beam size of $0.171\times0.155$", similarly to our program. We combined the calibrated measurements sets for both dataset using the CASA task CONCAT.% and created uv tables which we analyzed with GILDAS.
\end{itemize}

JWST NIRCam and MIRI images are publicly available for GS-9209 and for the ADF22-QGs. We reduced the data using the JWST Calibration Pipeline v1.19.1 \citep{Bushouse2025} with the CRDS pipeline mapping 1413. %During data reduction, 
We applied several custom-made modifications to the default pipeline to further improve the quality of the JWST mosaic \citep{Wang25, Sun26}. %We confirmed the accuracy of the ALMA and JWST alignment using the F444W image and extracted flux densities in the F770W and F2100W filters -- where contribution from warm or hot dust can be expected -- performing aperture photometry using Kron apertures with k=2.5 (using Source-Extractor v2.25.0).
Notably, QG2 is bright in F2100W, whereas QG1 and QG3 are detected at low signal-to-noise. 
%We use the flux densities from F356W and longer wavelengths to anchor the stellar component of our SED fitting, and the F2100W flux density to constrain the dust temperature incorporating our ALMA upper limits. More details can be found in Sect.~\ref{sec:sedfitting_qg2}.

\section{Reduction and analysis of ALMA data}\label{sec:analysis}

We calibrated ALMA observations with the standard CASA script. We split individual sources after binning the original frequency channels by $\times2$ and performed time averaging with 30s. We then exported {\it uvfits} files for further analysis in uv-plane with GILDAS\footnote{\url{https://www.iram.fr/IRAMFR/GILDAS/}}. 
We did not perform any cleaning of the data. In the case of GS-9209, the field is empty and thus it was unnecessary. In the case of the SSA22 field, we fitted and subtracted 6 bright star-forming sources in the protocluster, all detected in the [CII] line and continuum, to avoid contamination our targets' visibilities and potential biases in the uv-plane fit. In what follows we describe the analysis done on the spectra covering the expected frequency of the redshifted [CII] line. 
%We provide more details on the measurement and fitting of dust-continuum flux densities underneath [CII] in Sect.~\ref{sec:SED_analysis}.
We extracted the spectra of the quiescent galaxies using our custom pipeline scripts following the iterative procedure described in \cite{Daddi15}, see also \cite{Valentino2018, Valentino2020}. In the case of secure detections, we allowed fine tuning of the extraction position with small offsets from the nominal JWST positions, while for non-detections or weak signal (at the level of SNR $\gtrsim$3) we extracted at nominal positions. Spectra extracted over point sources are used when resolved profile extractions did not provide solid evidence for spatial extension in the sources, which eventually is the case only for M0138 and for QG3\footnote{QG3 has signs of a complex morphology in [CII], as discussed later.}. Given the resulting upper limits on the intrinsic ALMA sizes of the sources, fluxes might be underestimated by up to 10-20\% following this procedure \citep[see, e.g. ][]{Valentino2020}, depending on the (unknown) size of the sources. %In the case of GS-9209, where the data has very high spatial resolution, we extracted the spectrum using an exponential disk with a size of 0.2$''$, matching exactly the measurement at the longest wavelength MIRI data provided by \cite{Ji2024}, ensuring homogeneous comparison of ALMA and JWST/MIRI fluxes.

%\textbf{In order to maximize the frequency coverage of our observations the spectral setup was chosen  to bracket the [CII] line at the expect frequency with two spectral windows with an overlap of 0.2 GHz.} 

From the extracted spectra, our custom pipeline robustly estimates the continuum and its error, and searches for emission lines.  
For galaxies without evidence of the [CII] emission line, we measured the 3$\sigma$ integrated flux density upper limits;
%as the velocity integrated rms  \textbf{ measured over the spectral windows bracketing the expected line}, multiplied by the square root of the number of channels within the expected stellar FWHM and by the channel width in km s$^{-1}$; 
while in case of robust or tentative emission lines (SNR=3 or more), we adopted the line flux measurement from our standard pipeline, as discussed above, which finds the channel range maximizing the resulting SNR, which also defines the line full width at zero intensity (FWZI).
The integrated values are then converted into line luminosities using Eq. (1) in \citep{Solomon97} . Gas masses and upper limits are computed using  the \alphacii in Z18, thus $\rm{\alphacii^{18} = 31 \Msun/\Lsun}$. %We also quote  SFR$_{\rm [CII]}$ values or  upper limits  assuming the SFR-\Lcii relation from \cite{Schaerer2020}.
We also quote SFR$_{\rm [CII]}$ values by assuming from \cite{Schaerer2020}, which we use to gauge the position of our galaxies with respect to the star-forming MS and compare it to independent tracers.

\section{[CII] line and dust continuum measurements}
\label{sec:cii_results}
In what follows, we describe the peculiarities of each galaxy and we provide measurements for the integrated [CII] line fluxes and the underlying dust continuum flux densities, summarized in Table~\ref{table:tab_spec}. For M0138,   reported values are {not corrected for the magnification factor (unless explicitly specified)}. % $\mu$, that is $\mu$\Lcii, $\mu$\Mstar and $\mu \rm{M_{mol}}$.\\

\begin{table*}
\caption{Properties of our sources. Upper limits are given at 3$\sigma$. Values for M0138 are not corrected for the magnification factor and are flagged by an asterisk (*). For gas masses derived from [CII] we assume an \alphacii=31 \Msun/\Lsun. The line flux and luminosity of M0138 is scaled up by a factor of 1.2 to correct for point-source extraction (see Sect.~\ref{sec:cii_results}).}       
\label{table:tab_spec}
\centering 
\resizebox{\textwidth}{!}{\begin{tabular}{c c c c c c c c c c}   

\hline\hline         

ID & $\rm{\nu_{obs, [CII]}}$ & FWZI & \zspec$_{\rm{,\,opt}}$ & \zspec$_{\rm{,\,[CII]}}$ & $\rm{F_{[CII]}}$ & \Lcii & $\rm{SFR_{\cii}}$ & M$_{\rm{gas, [CII]}}$ & f$_{\rm{gas,[CII]}}$ \\ 
 & [GHz] & [km~s$^{-1}$] & & &[Jy~km~s$^{-1}$] & [$10^{8}$ \Lsun] & [\Msun $\rm{yr^{-1}}$] & [$10^{9}$ \Msun] & [\%]\\   
\hline                        \\         
M0138 & 644.62 & 837.6 & 1.9486 & $1.948^{+0.002}_{-0.002}$ & $3.02\pm0.72$ (*) & 3.83 $\pm$ 0.81 (*) & 35.7 $\pm$ 7.5 (*)& 11.86 $\pm$ 2.50 (*)& 0.19 $\pm$ 0.04 \\\\

QG1    & 464.74& 561.6 & 3.092$^{+0.008}_{-0.004}$ & 3.089$^{+0.004}_{-0.004}$ & $0.67\pm 0.16$ & 2.25 $\pm$ 0.55 & 20.9 $\pm$ 5.11 & 6.96 $\pm$ 1.70 & 5.4 $\pm$ 1.3 \\\\

QG2    & 465.23& 624.7 & 3.0851$^{+0.0001}_{-0.0001}$ & $3.0851^{+0.004}_{-0.004}$ & $1.36\pm 0.18$ & 4.55 $\pm$ 0.60 & 42.4 $\pm$ 1.1 & 14.09 $\pm$ 1.87  & 25.6 $\pm$ 3.4 \\\\
QG3    & 463.93 & 161.66 & 3.0744$^{+0.0003}_{-0.0003}$ & $3.0965^{+0.0007}_{-0.0007}$ & $0.19 \pm 0.03$ & 0.64$\pm$0.10 & 6.0 $\pm$ 1.2 & 2.0 $\pm$ 0.3 & 2.8 $\pm$ 0.4\\\\  
GS-9209 & -      &   -   & 4.6582$^{+0.0002}_{-0.0002}$ & - & $<0.0356$ & $<$ 0.69& $<$ 6.4 & $<$ 2.1 & $<$5.2  \\\\      
\hline   %inserts single line

\end{tabular}}
\end{table*}

\textbf{M0138:} its spectrum contains an emission line at $\rm{\nu_{obs}=644.615 \,GHz}$ at 4.7 $\sigma$ for which we measure $\rm{FWZI = 838 \, km \,s^{-1}}$ and $\rm{z_{[CII]}=1.948\pm0.002}$ in good agreement with the stellar velocity dispersion and redshift, respectively, from rest optical spectra in \cite{Newman18} and \cite{Newman2025}. Because M0138's morphology is extended and its image is significantly stretched, the line flux and its underlying continuum measured after point-source extraction have been scaled up by a factor of 1.2, that is the ratio between the point-like extraction (that we use) and the lensing model-based extraction in \cite{Gobat22}. We thus obtain an integrated line flux $\rm{F_{[CII]}=3.02 \pm 0.72 \, Jy \, km\, s^{-1}}$, and a flux density of the dust continuum underneath [CII] of $\rm{F_{cont}=  2.38 \pm 0.45 \, mJy}$. This gives us $\rm{L_{[CII]} = (3.83 \pm 0.81) \times 10^{8} \, \Lsun}$, $SFR_{\rm{[CII]}}=35.7 \pm 7.5$ \Msun yr$^{-1}$. This latter value is in agreement with the magnified SFR from Image 1 in \cite{Newman18} ($\rm{31 \pm 9\, \Msun\,yr^{-1}}$) and suggests that the gas seen by [CII] might be involved in residual star formation (about 2.6 \Msun $yr^{-1}$ de-magnified). Lastly, we obtain $\rm{M_{mol}=(11.86\pm 2.50)\times 10^{9} \,\Msun}$ and $\rm{f_{mol}=0.19\pm 0.04}$\% confirming the extreme gas depletion reported for this galaxy by \cite{Whitaker21a}. We note that our conclusions on \fg or on $\rm{\delta_{GD}}$ are independent of the magnification factor. 

\begin{figure*}[]
 \centering
    \includegraphics[width=0.9\textwidth]{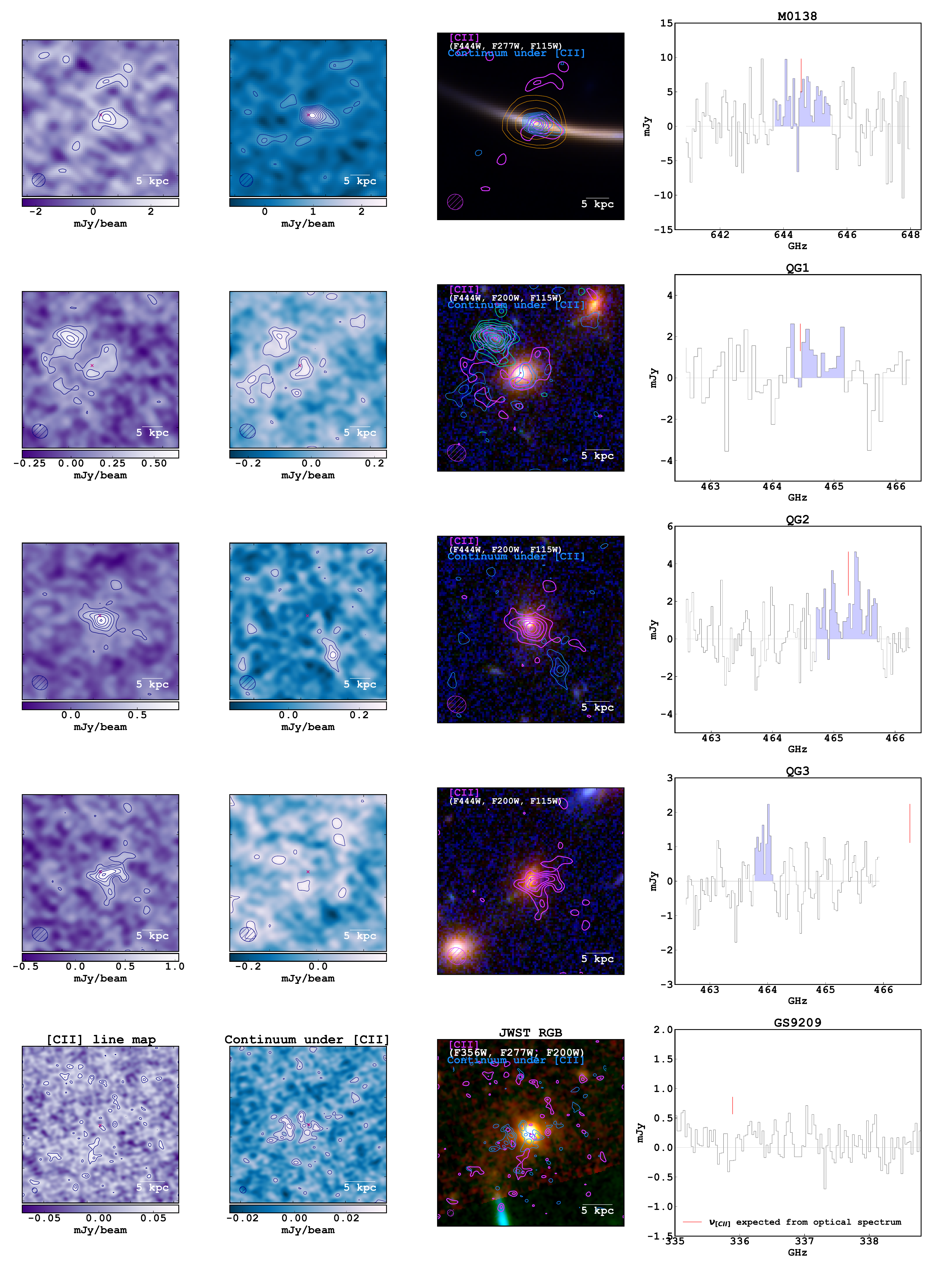}
    \caption{Columns from left to right: [CII] line emission; dust continuum underneath [CII]; [CII] and dust continuum contours (pink and blue, respectively) overlayed on JWST RGB images; [CII] spectra, binned for clarity.  Contours are shown from 2 to 10 $\sigma$ in steps of 1$\sigma$. Green contours in the RGB image of QG1 mark the [CII] emission of a satellite galaxy, QG1-Sat (see Sect.~\ref{sec:cii_results}). Orange contours for M0138 show the dust continuum detection in \cite{Whitaker21a}. Red vertical lines in the spectra mark the expected observed-frame frequency of [CII] based on the spectroscopic redshift derived from optical spectra. We note that the poor coverage of the expected [CII] frequency for QG3 is due to the ALMA pointing and spectral setup being optimized for QG1. In our dataset we detect a line at the position of QG3 in agreement with the redshift distribution of the galaxy group.} 
    \label{fig:all}
\end{figure*}

In Fig.~\ref{fig:all} we overlay the [CII] contours and its continuum to the JWST RGB image of M0138 (F444W, F277W, F115W, program GO \#06549, PI: J. Pierel). The composite image shows that the galaxy is composed by a red core and an extended bluer disk \citep{Newman2025} emitting at rest $\sim$ 3900 \AA, distributed over the whole length of the arc connecting Image 1 and Image 2. The [CII] and its continuum are co-spatial and located at the position of the bulge, consistently with the position of the 1.3 mm detection. Our contours show that the signal is marginally extended in the direction of the stretch.
MO138 is also detected in dust continuum underneath [CII], with a flux density of $\rm{F_{cont}=  2.38 \pm 0.45 \, mJy}$ (already scaled) that we use in Sect.~\ref{sec:SED_analysis} to further constrain the dust mass. %In Fig.~\ref{fig:all} we overlay the [CII] contours and its continuum to the JWST RGB image of M0138. 
%The composite image shows that the galaxy is composed by a red core and an extended bluer disk \citep{Newman2025} emitting at rest $\sim$ 3900 \AA, distributed over the whole length of the arc connecting Image 1 and Image 2. The [CII] and its continuum are co-spatial and located at the position of the bulge, consistently with the position of the 1.3 mm detection. Our contours show that the signal is marginally extended in the direction of the stretch. 

\textbf{QG1:} its spectrum extracted at fixed JWST position shows no emission line and no continuum detection, yielding a 3$\sigma$ upper limit on the integrated line flux of $S_{\rm{[CII]}}<0.10$ Jy km s$^{-1}$, which gives $L_{\rm{[CII]}}<0.33\cdot 10^8 \, L_{\odot}$. However, we detect plausible signal both in [CII] and in dust continuum, 0.35" ($\Delta \alpha=0.33", \Delta\delta=0.05"$) offset from the nominal position of the source. The two components are co-spatial and they also overlap with stellar residuals detected in the F444W image of the galaxy, which will be discussed in more detail in Sect.~\ref{sec:mergers}. Extracting the spectrum at this position reveals a tentative line falling at $\rm{\nu_{obs}=464.742 \,GHz}$ at 4.1 $\sigma$, with $\rm{FWZI_{[CII]} = 562 \pm 278 \, km \,s^{-1}}$. This yields $\rm{z_{[CII]}=3.089\pm0.004}$, in agreement with the rest optical redshift from \cite{Kubo21} and with the CO(3-2) redshift from \cite{Umehata2025}. The underlying dust continuum  is also marginally detected with $\rm{F_{cont}= 0.436 \pm 0.142 \, mJy}$. Although such results will need to be confirmed with deeper datasets, we consider the spatial overlap of the three independent galaxy components (stellar, [CII], dust continuum) and the redshift agreement between the [CII] line, the optical and CO(3-2) spectra as reasonable indications that emission is present. The line width is in agreement with the binned CO(3-2) line presented in U25 ($\rm{FWZI_{CO(3-2)} = 600 \pm 254 \, km \,s^{-1}}$), and broadly consistent with the rest optical velocity dispersion of QG1 ($\rm{FWHM{_\star}\sim750}\,km\,s^{-1}$).
At this latter position we measure $\rm{S_{[CII]}=0.67 \pm 0.16 \, Jy \, km\, s^{-1}}$, which corresponds to $\rm{L_{[CII]} = 2.25 \pm 0.55 \cdot 10^{8} \, \Lsun}$, $\rm{M_{mol} = 6.96 \pm 1.70  \cdot 10^{9} \, \Msun}$ and $f_{\rm{mol}} = 5.4 \pm 1.3$\%. Alternately, \Lcii gives us $SFR_{\rm{[CII]}}=20.9 \pm 5.1$ \Msun yr$^{-1}$, at the level of the $3\sigma$ upper limit quoted in \cite{Kubo21}.
Additionally, we detect with very high confidence a second [CII] emission line 2" offset from QG1 but at the same redshift \zspec=3.0887$\pm$0.0001. This detection is centered on a star-forming satellite galaxy (hereafter QG1-Sat) next to QG1 and visible in composite JWST/NIRCam images (see Fig.1, green contours). QG1-Sat shows a [CII] line luminosity of $L_{\rm{[CII]}}=1.47 \pm 0.12 \cdot 10^8 \, L_{\odot}$ and $M_{\rm{mol}}=4.56 \pm 0.37\cdot 10^9 \, M_{\odot}$ and is well within the beam of the CO(3-2) observations\footnote{The WCS uncertainty of the NIRCam mosaics is 0.01" for RA and 0.02" for Dec (smaller than the pixel size 0.03") and the positional uncertainty for the [CII] line from QG1-Sat is $\sim$0.12".}. The line has a $\rm{FWZI_{[CII]} \sim 150\, km \,s^{-1}}$, thus much narrower than [CII] and CO(3-2). Although we can expect that both sources contribute to the CO(3-2) emission line, the match of the [CII] and CO(3-2) FWZI of QG1 tends to suggests that the CO line originates from QG1 primarily. Additionally, mass-dependent variations in $\rm{\alpha_{CO}}$ \citep[e.g.,][]{Schruba2012} may potentially produce a fainter CO line for the satellite ($\sim30$ times less luminous in F444W), despite the two galaxies having similar [CII] luminosities. Higher resolution CO observations and/or a deeper [CII] spectrum focused on QG1 are needed to assess how much of the CO luminosity measured in U25 can be ascribed to QG1, revising its gas fraction accordingly. Hereafter, we will assume that the CO(3-2) line originates from QG1 and we will discuss implications for the \alphacii in Sect.~\ref{sec:alphacii_cal}.
Interestingly, QG1-Sat is separated from QG1 by only $\sim$50 km s$^{-1}$ in velocity space and, as already mentioned, is about $\sim30$ times less luminous than QG1 in F444W. Hence it is very likely a lower mass satellite not detected in the previous ground-based K-band images \citep[e.g.][]{Kubo16}. QG1-Sat has a disturbed morphology, with a red component plus a bluer, clumpy stellar tail visible in multiple JWST bands, including MIRI/F770W. Judging from the pair configuration and from QG1's stellar residuals shown in Fig.~\ref{fig:residuals}, the two galaxies may have had an interaction in their recent past, resulting in an asymmetric morphology for QG1 and in an ejected stellar stream for QG1-Sat.

\textbf{QG2:} its spectrum shows an emission line at $\nu_{\rm{obs}}$=465.23 GHz corresponding to \zspec=3.0851$\pm$0.0007, in good agreement with its [OIII]-based redshift. We measure a line width of FWZI = 624.65 $\pm$ 225 km s$^{-1}$. 
The integrated flux density is $S_{\rm{[CII]}}=1.36\pm0.18$ Jy km s$^{-1}$ (7.7$\sigma$), which gives us $L_{\rm{[CII]}}=(4.55\pm0.60)\cdot 10^8 \, L_{\odot}$ and $SFR_{\rm{[CII]}}=42.4 \pm 1.1$ \Msun yr$^{-1}$. Its gas mass and fraction come to  
$M_{\rm{mol}}=(14.1\pm 1.9)\cdot 10^9 \, M_{\odot}$ and $f_{\rm{mol}}=25.6\pm3.4$\%.

Although the optical [OIII] line shows the presence of AGN-driven outflows, we do not have sufficient SNR to detect broad line wings. We note, however, that the peak position of the [CII] emission is offset by 0.14" from the peak of the stellar component of QG2, which is $2\sigma$ significant given  the positional uncertainty of the image computed as $\rm{FWHM_{BEAM}/(2.35\times{\sqrt{SNR}}})$. A careful inspection of the available JWST/NIRCam images reveals that the galaxy has an extended halo (2.3"$\times$1.3") in the F200W, F356W and F444W filters. We modeled the central component of QG2 in the F444W filter with an empirical PSF taken from the same image. The residuals shown in Fig.~\ref{fig:residuals} reveal a lopsided disk component, likely the result of past galaxy merger which might have triggered the central AGN and outflow. Since the [CII] peak is offset along the same axis as the disk, we speculate it to be real and likely physically related to such interaction (see more discussion in Sect.~\ref{sec:mergers}). Dust continuum is not detected.

\textbf{QG3}: its spectrum shows a weak but fairly significant ($6\sigma$) emission line at $\nu_{\rm{obs}}$=463.93 GHz, corresponding to \zspec=3.0965$\pm$0.0003, in disagreement with the [OIII]-based redshift of QG3 (see Table~\ref{table:tab_spec}) but in better agreement with the redshift distribution of the AzTEC14 galaxy group. We note that the profile is fairly narrow (FWHM=162$\pm$60 km s$^{-1}$) to be associated to a massive galaxy such as QG3.
Taking the results at face value, the integrated flux density is $S_{\rm{[CII]}}=0.19\pm0.03$ Jy km s$^{-1}$, which gives us $L_{\rm{[CII]}}=(0.64\pm0.10)\cdot 10^8 \, L_{\odot}$ and $SFR_{\rm{[CII]}}=6.0 \pm 1.2$ \Msun yr$^{-1}$. Its gas mass and gas fraction are 
$M_{\rm{mol}}=(2.0\pm0.3)\cdot 10^9 \, M_{\odot}$ and $f_{\rm{mol}}=2.8\pm0.4$\%, respectively. Dust continuum is not detected.

\textbf{GS-9209:} its spectrum contains no detectable emission line, as shown in Fig.~\ref{fig:all}. The 3$\sigma$ upper limit on the integrated flux density is $S_{\rm{[CII]}}<$0.0356 Jy km s$^{-1}$, which converts into $L_{\rm{[CII]}}<0.69\cdot 10^8 \, L_{\odot}$, $M_{\rm{mol}}<2.14\cdot 10^9 \, M_{\odot}$ and $f_{\rm{mol}}<5.2$\%. Alternatively, \Lcii gives a 3$\sigma$ SFR upper limit of $SFR_{\rm{[CII]}}<6.4$ \Msun yr$^{-1}$, about 3 times higher the SFR measured from the narrow \Ha component of the NIRSpec spectrum in \cite{Carnall23}.
We detect significant dust continuum emission from this galaxy, with $\rm{F_{cont}=0.064\pm0.017\, mJy}$, extracting signal at fixed position and using the same shape and size of the MIRI 21 $\mu m$ emission from \cite{Ji2024} (an exponential disk with a size of 0.2$''$), to enable the study of its SED in Sect.~\ref{sec:gs_sed}. As discussed later, there is evidence for additional ALMA flux on wider scales. 

\section{Constraints on the dust temperature}\label{sec:SED_analysis}

\begin{figure}[]
 \centering
    \includegraphics[width=\columnwidth]{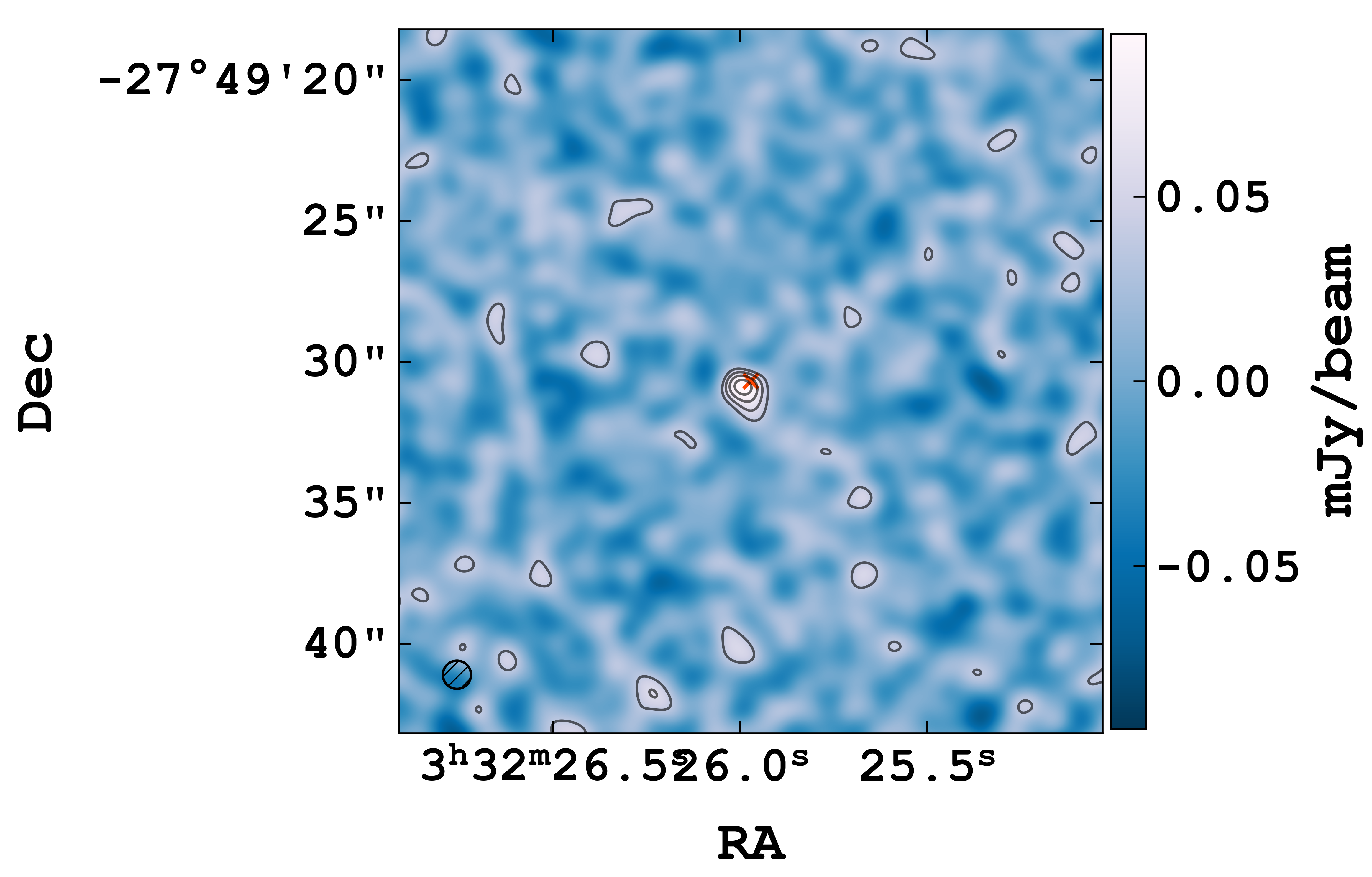}%
    % \caption{2 Figures side by side}%%
    \caption{Dust continuum emission from GS-9209 tapered with a 2D gaussian and uv=100 weights, yielding a beam of approximately 1". The red cross marks the brightest pixel of the stellar component in the F560W filter. Contours are drawn at 2,3,4,5$\sigma$.}
    \label{fig:gs_tapered}
\end{figure}

Motivated by the recent reports of diffuse, hot dust ($\rm{T_{d}\sim 47 \,K }$) on scales 2-3 times larger than the rest-optical stellar component of GS-9209 by \cite{Ji2024} we here attempt to constrain the MIR-to-FIR SED of our sources based on weak continuum detections underneath the \cii line (or their upper limits), adding information from MIRI imaging at 7.7$\rm{\mu m}$ and 21$\rm{\mu m}$ and archival mm and radio measurements whenever available. We thus attempt a measure of their total infrared luminosities and dust temperatures.

We perform SED fitting using MICHI2 \citep{Liu2021}, which does not require imposing energy balance between the optical reddening of the stellar component and the FIR dust component, as expected for quenched sources. We adopt: BC03 stellar templates \citep{bc03} to only model fluxes at $\rm{\lambda_{rest}>1 \mu m}$ as a baseline upon which dust emission is recovered; a Mid-IR AGN component based on observationally calibrated AGN torus templates from \cite{Mullaney2011}; dust templates from \cite{DL07} (hereafter DL07) spanning a large range of effective dust temperatures, where dust grains are exposed to a PDR whose strength of the interstellar radiation field intensity ranges from U$_{\rm{min}}$ to U$_{\rm{max}}$=10$^7$; and, for fits incorporating a radio constraint, we use a default radio power-law component tied to the total infrared luminosity through the infrared-radio correlation, using a mass- and redshift-dependent q$_{\rm{IR}}$ from \cite{Delvecchio21}. The latter point is relevant only for QG2, which is detected in radio, while upper limits for the other sources are not stringent enough to constrain the fits in the radio regime. 
%Due to the limited star formation histories implemented in MICHI2, we refrain from fitting the rest-frame optical stellar component of our targets.
We convert $\rm{<U>}$ into $T_{d}$ using the relation $\rm{<U> = (T_{d}/18.9\,K)^{6.04}}$ resulting from \cite{magdis12}, following \cite{Jin22}.
All results can be found in Table~\ref{table:SED_analysis}, where, for completeness, we include outputs obtained with and without including an AGN torus in the fit.

\begin{table*}
\caption{Dust continuum results. Upper limits are given at 3$\sigma$ and in mJy/beam. 
%Fits with MICHI2 are taken using BC03 templates with a constant SFH  with multiple ages and a $\rm{q_{IR}}$ fixed to the \cite{Delvecchio21} relation. 
M$_{\rm{d,[CII]}}$ is computed assuming a standard $\delta_{GD}$=100. Whenever we obtain a lower limit for $\rm{T_{d}}$, $\rm{M_{d}}$ should be considered an upper limit.}

\label{table:SED_analysis}     
\centering   
\resizebox{\textwidth}{!}{\begin{tabular}{c c c c c c c c } 
\hline\hline         

ID & F$_{\rm{cont}}$ [mJy] & log($\rm{L_{IR,TOT}}$/\Lsun) &  log($\rm{L_{IR,repr}}$/\Lsun)&  log($\rm{M_{d,TOT}}$/\Msun) & <U> & $\rm{T_{d}}$ [K] &  log(M$_{\rm{d,[CII]}}/\Msun$) \\
\hline                        \\        

M0138 (no AGN)& 2.38$\pm$0.46 & 11.62 $\pm$ 0.36  & - & 8.74 $\pm$ 0.68 & 25.6$\pm$24.4 & 32.33$^{+3.79
}_{-12.85}$  & 8.1 $\pm$0.10   \\\\
\hline \\

QG1 (w AGN)& 0.436 $\pm$ 0.142 & 10.86 $\pm$ 0.20 & 10.96 $\pm$ 0.09 & 6.81 $\pm$ 0.51 & 36.25$\pm$16.25 & 34.3$^{+2.2}_{-3.2}$  & 7.8$\pm$ 0.1   \\\\
QG1 (no  AGN)& " & 10.41 $\pm$ 0.71 & " & 6.53 $\pm$ 0.82 & 40.54$\pm$23.54 & 34.9$^{+2.7}_{-4.7}$  & "   \\\\

\hline \\ 
QG2 (w AGN)& $<$ 0.49 & 12.34 $\pm$ 0.09& 10.69 $\pm$ 0.22 &$<$ 7.45 & $>$ 497.17 & $>$52.8 & 8.1 $\pm$ 0.1  \\\\

QG2 (no AGN)& " & 12.30 $\pm$ 0.06& " &$<$ 7.45 & $>$ 497.17 & $>$52.8  & "  \\\\
\hline  \\
QG3 (w AGN)& $<$ 0.146 & $<$11.36 & - &$<$ 7.11 & $>$ 2.6 & $>$22.14  & 7.3 $\pm$ 0.14  \\\\

QG3 (no AGN)& " & $<$11.39 & - &$<$ 7.63 & $>$ 25.1 & $>$32.2  & "  \\\\
\hline  \\

GS-9209 (w AGN) & 0.064$\pm$0.017 & 11.12 $\pm$ 0.66 & 9.78$^{+0.37}_{-0.61}$ & 6.98 $\pm$ 0.91 & 245.4 $\pm$ 241.9 & $\rm{47.0 ^{+5.7}_{-23.7}}$ & $<$7.3 \\\\
GS-9209 (no AGN) & " & 11.78 $\pm$ 0.18 & " &$<$ 7.45 & $>$ 163.2 & $\rm{>43.9}$ & " \\\\

\hline  
\end{tabular}}
\end{table*}

\subsection{Evidence for warm dust over a diffuse and extended stellar halo around GS-9209}\label{sec:gs_sed}

The ALMA cube shows a weak and extended dust continuum signal at the JWST position of the source and corresponding to an extended, faint stellar halo visible in the F200W, F356W and F444W filters. 
We perform flux extraction in the {\it uv} space at the JWST position of the galaxy using a Sérsic profile with n=1 corresponding to r$_{\rm{eff}}$=1.8$\pm$0.5 kpc measured in \cite{Ji2024}, thus not including the leftmost continuum emission visible in Fig.~\ref{fig:all}. Thus, we make sure not to include emission offset from the main stellar component of the galaxy. 
Although a fit with n=1 might not fully grasp the exact profile of the galaxy, it has the advantage of downplaying the flux contribution from the central AGN-dominated component, as noted in \cite{Ji2024}, and extending further out of the stellar component of the galaxy.
We obtain $F_{\rm{cont}}=0.064\pm0.017$ mJy, which we add to the FIR SED collected in \cite{Ji2024} (see their Fig.7) which refers to the integrated stellar profile of the galaxy (r$<$0.7"). We also incorporate their 1.2 mm dust continuum upper limit in our SED fitting.
 
The best fitting model (without an AGN component) is shown as a black solid curve in Fig.~\ref{fig:gs_sed_noagn} (see Appendix~\ref{sec:michi2_plots}) and it is largely dominated by a warm dust component with $T_d>43.9$ K, whereas restricting the fit to a single cold dust component or to a combination of cold dust and AGN components does not fit the available JWST/MIRI photometry equally well. The associated dust mass upper limit is fairly uncertain with $\rm{log(M_{\rm{d, TOT}}/\Msun)<7.5}$. 
Assuming a standard gas-to-dust ratio $\delta_{GD}$=100, we obtain a cold gas mass of  log($M_{\rm{gas}}/\Msun)<9.5$, which would result in \Lcii$<0.9 \times 10^{8}$ \Lsun, consistent with our non detection. Our results support the presence of diffuse warm dust in GS-9209 as reported in \cite{Ji2024}. 

In order to evaluate if additional dust is present beyond what recovered based on the JWST profile, we measured the dust continuum signal-to-noise at increasingly large FWHM. We detect a peak of SNR=5.05 at FWHM=0.4" corresponding to a flux density of $f_{\rm{cont, [CII]}}=0.184\pm0.028$ mJy, about 2.8 times the flux density measured at FWHM=0.2". The excess signal seem to pertain to the diffuse stellar halo seen in the residual images from JWST. In Fig.~\ref{fig:gs_tapered} and Fig.~\ref{fig:residuals} (top left panel) we show a tapered image of GS-9209 with a synthesized beam of 1" (black contours and white dashed contours, in the two respective figures). Overall, we report an offset of 0.35" between the centroid of the tapered image and the nominal position of the galaxy from the JWST/MIRI F560W filter.

\subsection{Evidence for warm dust also for QG2}
\label{sec:sedfitting_qg2}

We extracted JWST/MIRI F770W and F2100W fluxes using automatic apertures and we incorporated them to the SED of \cite{Kubo22}, together with our rest 158$\rm{\mu m }$ dust continuum upper limit. We fit QG2 anchoring the normalization of the stellar template to the points between 4.8 and 7.7$\rm{\mu m }$ (1.1--1.8$\rm{\mu m }$ rest). We fixed the $q_{IR}$ using the \cite{Delvecchio21} relation and we impose E(B-V)$<$0.12, taking the maximum value allowed from \cite{Kubo22}. For QG2, the best fits systematically hit the upper boundary on <U>, which arises from the maximum U$_{\rm{min}}$ allowed in the DL07 models. Therefore, we quote the lower uncertainty on our best fit value as a lower limit on <U> and on $\rm{T_{d}}$, thus $\rm{<U> > 497.17}$ and $\rm{T_{d}>52.8}$K. The artificial upper boundary on $\rm{T_{d}}$ implies that the lower uncertainty on dust masses could be lower. Therefore, whenever we have a lower limit on $\rm{T_{d}}$, dust masses should be regarded as upper limits. 
These results do not change if we force a substantial AGN torus component to the MIR. However, we note that the F2100W emission in QG2 is resolved and consistent with a Sérsic profile. %(see Fig.~\ref{fig:QG2_MIRI})
%, leaving no residuals after the fit. 
This suggests that an AGN torus does not contribute much to MIR fluxes, and that QG2 hosts dust (and likely gas) at larger temperatures compared to expectations from the size evolution of quiescent galaxies described in \citep{Magdis21} and larger than in star-forming galaxies at the same redshift (see Fig.~\ref{fig:tdust_evo}).

As can be seen in Table~\ref{table:SED_analysis} QG2 has very high LIR and $\rm{T_{d}}$ values. They stem from including the radio fluxes in the SED fitting while imposing a $q_{\rm IR}=2.45$, so that MICHI2 selects the largest LIR that still allows it to properly fit the normalization of the radio component. In this case we see a mild radio excess, which would imply $q_{\rm IR}<2.30$, within the scatter of the \cite{Delvecchio21} relation, thus not formally requiring any radio AGN component in the galaxy. We do not find a low $q_{\rm IR}=0.61$ described in \cite{Kubo22} likely due to our larger LIR upper limit found including the MIRI detection to our fits.
On the other hand, if we remove our constraints on $q_{\rm IR}$ all parameters become unconstrained when including the AGN torus component in the fit. As mentioned above, however, the rest-frame 3-4 $\mu m $ emission is resolved and does not show evidence of a dominant central source. Removing the AGN torus template \emph{and} the constraint on $q_{\rm IR}$ still yields $\rm{log(LIR/\Lsun)= 12.13\pm0.07}$ and $\rm{T_{d}> 49.8\, K}$.

\subsection{Constraints on $T_{d}$ and $M_{d}$ for M0138}
\label{sec:m0138_sed}
For this galaxy we do not include an AGN torus in our fit since it has no evidence of radiatively efficient accretion from a SMBH and it is also undetected in X-rays \citep{Newman2025}. We note that including an AGN component in our fit would not substantially affect the results as we only have submm-measurements. 
Using our rest 158 $\mu m$ dust continuum measurement and including the 1.3 mm flux density from \citep{Gobat22} we find only weak constraints on $\rm{T_{d}}$ and $\rm{M_{d}}$, with $\rm{T_{d}=32.3^{+3.8}_{-12.8}}$ and $\rm{log(M_{d}/\Msun)=8.7}$. This latter value is about 2.5 smaller than in \cite{Gobat22}. We recall that \cite{Gobat22} derived $\rm{M_{d}}$ assuming $\rm{T_{d}=21\,K}$.
Our results support a somewhat warmer dust but are also still consistent with that assumption.

\subsection{Constraints on $T_{d}$ for QG1 and QG3}

We used MIRI imaging  at F2100W, where both sources are fairly weak. %consistent with emission by the stellar component. 
For QG1 the ALMA continuum at 3$\sigma$ and the MIRI constraint result in $T_{\rm d}\sim34\,K$, warmer than expected for quenched galaxies and more similar to star-forming galaxies at the same redshift, but less extreme than for GS-9209 and QG2. 
This may suggest that gas has been primarily consumed through secular evolution for this galaxy, without extra heating from other sources.
The constraints on LIR and $\rm{M_{d}}$ tend to be rather dependent on whether we do or do not include an AGN torus component in the fit.
As for QG3, lacking any ALMA detection, our fits only yield lower limits to $\rm{T_{d}}$ which are not constraining.you

\section{Discussion}\label{sec:discussion}

\subsection{Constraints on $\alphacii$ and implications for gas mass measurements for high-z QGs}
\label{sec:alphacii_cal}

In this section, we attempt to constrain \alphacii using M0138 and QG1, the only two targets in our sample with detections in both [CII] and an independent gas mass tracer. Although based on just two galaxies, this represents, to our knowledge, the first attempt to constrain \alphacii in systems with strongly suppressed star formation activity, particularly at cosmic noon. We begin with M0138 and then consider QG1, leveraging detections in either dust continuum or CO(3–2). The accuracy of \alphacii depends on the adopted conversion factors and their uncertainties (\dgr\ or \alphaco). Given the limited sample size, we do not attempt to quantify the intrinsic scatter in \alphacii and instead adopt standard conversion factors without accounting for their dispersion, as is commonly done in the literature.

\subsubsection{Clues from dust continuum in M0138}

We constrain \alphacii for M0138 using the most accurate dust mass and stellar metallicity measurements currently available. To derive the gas mass, we estimate \dgr\ from the gas-phase metallicity-dependent scaling relation of \cite{magdis12}. Although no gas-phase metallicity is available for M0138, spatially resolved stellar metallicity measurements for Image~1 from \cite{Jafariyazani20} provide $\rm{[Fe/H]=0.21\pm0.06}$ and $\rm{[Mg/Fe]=0.51\pm0.02}$ in the core region, where dust is detected. We conservatively approximate [O/H]$\approx$[Fe/H]+[Mg/Fe]\footnote{We conservatively ignore that gas metallicity may exceed the stellar value and that O is more abundant than Mg.}, yielding $\rm{12+log(O/H)=9.4\pm0.1}$ and a low $\rm{\dgr=17}$. This value is $\sim$5 times lower than the commonly adopted \dgr=100 and significantly below predictions from recent simulations \citep[e.g.,][]{Whitaker21a}, which invoke efficient dust destruction. Adopting the \cite{magdis12} scaling implicitly assumes negligible dust destruction.
Having derived the \dgr\ for this galaxy, we proceed to convert the dust mass obtained by our SED fitting, as well as the dust mass previously obtained in \cite{Gobat22}, into gas masses that we use for our test.
Using the dust mass from Sect.~\ref{sec:m0138_sed}, we derive $\rm{\mu M_{mol}=9.3_{-7.4}^{+35.4}\times10^{9}\,\Msun}$ and $\alpha_{[CII]}=24_{-20}^{+125}\,\Msun/\Lsun$, with uncertainties too large for a meaningful constraint. Conversely, adopting the dust mass from \cite{Gobat22}, based on a single cold dust component, yields $\rm{\mu M_{mol}=(2.21\pm0.04)\times10^{10}\,\Msun}$ and $\alpha_{[CII]}=87\pm12\,\Msun/\Lsun$, a factor of 2.8 higher than in Z18. Thus, we find rough agreement with $\alphacii^{Z18}$ and its scatter when adopting a low \dgr\ motivated by the stellar metallicity. This may indicate either inefficient dust destruction or compensation by internal dust production mechanisms. We note that, given the unusually high stellar metallicity of M0138, larger samples of quenched galaxies at similar redshifts are required to better constrain \alphacii in these sources.

\subsubsection{Clues from CO(3-2) in QG1: non-evolving [CII]/CO ratio}
We take advantage CO(3-2) detection of QG1 from U25 to attempt a first estimate of \alphacii\ in quiescent galaxies at $z=3$. %We make use of the excitation ratio $\rm{r_{31} = 0.5}$ to convert from $\rm{L_{CO(3-2)}}$ to $\rm{L_{CO(1-0)}}$ \citep{Daddi15}.

Assuming $\rm{\alphaco=4.4\,\Msun/(K\,km\,s^{-1}\,pc^{-2})}$ and attributing all CO(3–2) emission to QG1 yields $\alphacii=81\pm 24\,\Msun/\Lsun$, a factor $\sim2.6$ above $\alphacii^{Z18}$ and consistent with the trend in Z18 of  \alphacii with decreasing SSFR relative to the main sequence at SSFR/SSFR(MS)$<$0.3. Conversely, adopting $\rm{\alphaco=0.8\,\Msun/(K\,km\,s^{-1}\,pc^{-2})}$ gives $\alphacii=14.7\pm4.3\,\Msun/\Lsun$, within scatter of the of the Z18 relation and about a factor of two lower. Overall, the canonical \alphacii\ remains broadly consistent within the uncertainties. Higher quality data are needed to assess the potential blending of the CO(3-2) line with QG1-Sat.

\begin{figure*}[]
 \centering
    \includegraphics[width=0.8\textwidth]{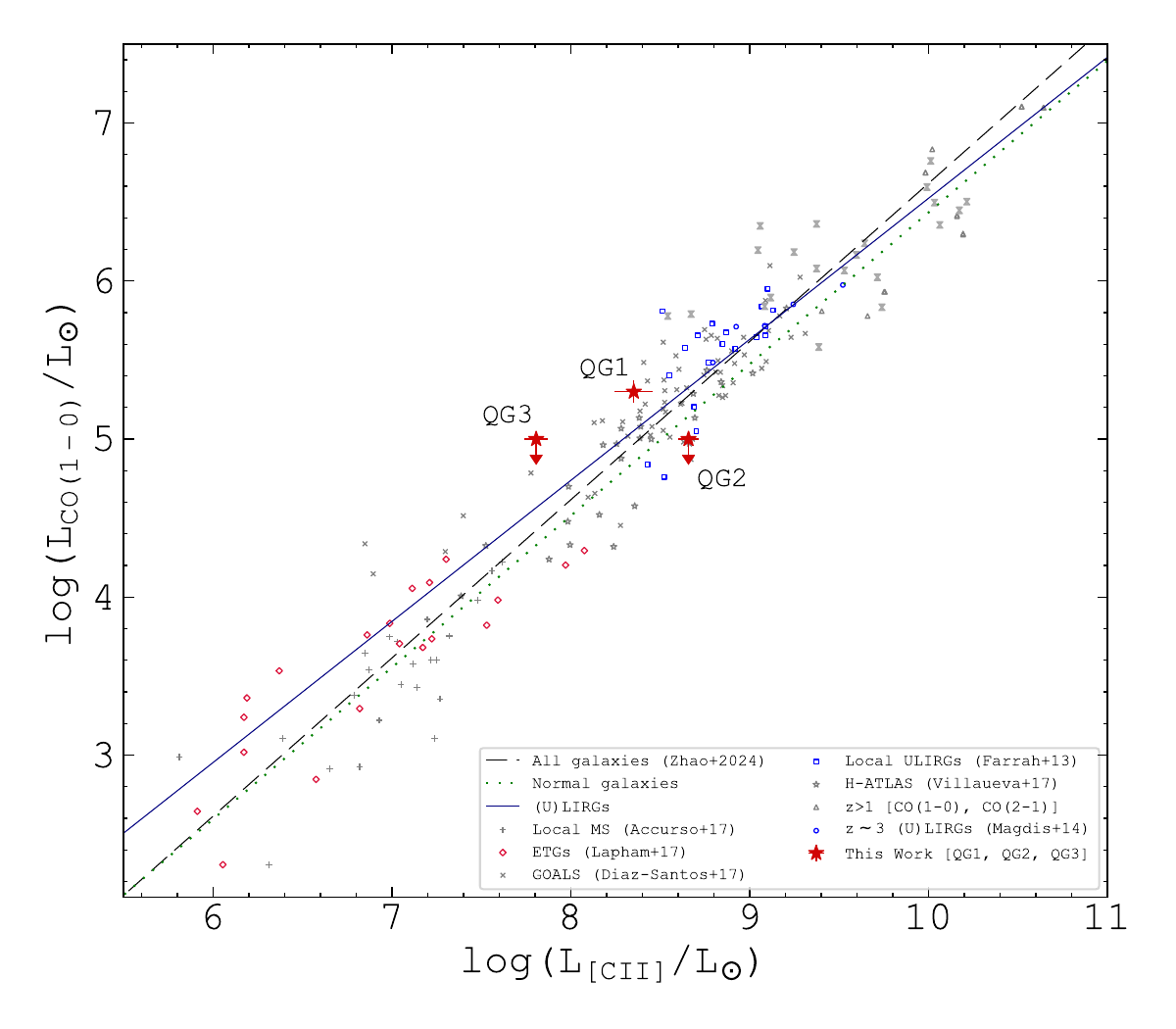}%
    \caption{Relation between the $\rm{L_{CO(1-0)}}$ and \Lcii. Red stars mark the line luminosities (or upper limits) for QG1, QG2 and QG3. Fitted curves show the relation for all galaxies considered in \cite{Zhao24} (dashed black), normal galaxies including ETGs (dotted green) and (U)LIRGS (solid blue). Blue and red empty markers highlight the position of ETGs and (U)LIRGS to ease the comparison with our sources. Plot adapted from \cite{Zhao24}.} 
    \label{fig:Lcii_Lco}
\end{figure*}

Using an excitation ratio $\rm{r_{31} = 0.5}$ to convert from $\rm{L_{CO(3-2)}}$ to $\rm{L_{CO(1-0)}}$ \citep{Daddi15}, the inferred $\rm{L_{CO(1-0)}}$ and \Lcii place QG1 on the local relation of \cite{Lapham2017, Zhao24}, alongside early-type and star-forming galaxies (see Fig.~\ref{fig:Lcii_Lco}). Interestingly, QG1 lies on the high luminosity range of the ETG distribution, approaching ULIRG-like values, while upper limits for the other ADF22-QGs are consistent with the local scatter. Although this relation does not tightly constrain absolute conversion factors, it suggests that the $\alphacii/\alphaco$ ratio remains approximately invariant from $z=0$ to $z=3$ (given a scatter $\sim0.3$ dex), and that \alphacii\ may not strongly deviate from standard values in $z=3$ ETGs, provided that \alphaco\ is stable.
%The $\rm{L_{CO(1-0)}}$ of QG1 may be overestimated if QG1-Sat contributes to the CO(3–2) emission. However, despite comparable [CII] luminosities, QG1-Sat is $>30\times$ less massive judging from the the F444W flux ratio, pointing to a significantly higher \alphaco\ and a negligible CO contribution. This supports associating the CO(3–2) emission primarily with QG1, consistent with \cite{Umehata2025}.

\subsection{Widespread merger signatures underlying [CII] and dust continuum offsets}
\label{sec:mergers}

\begin{figure}[h!]
 \centering
    \includegraphics[width=1.05\columnwidth]{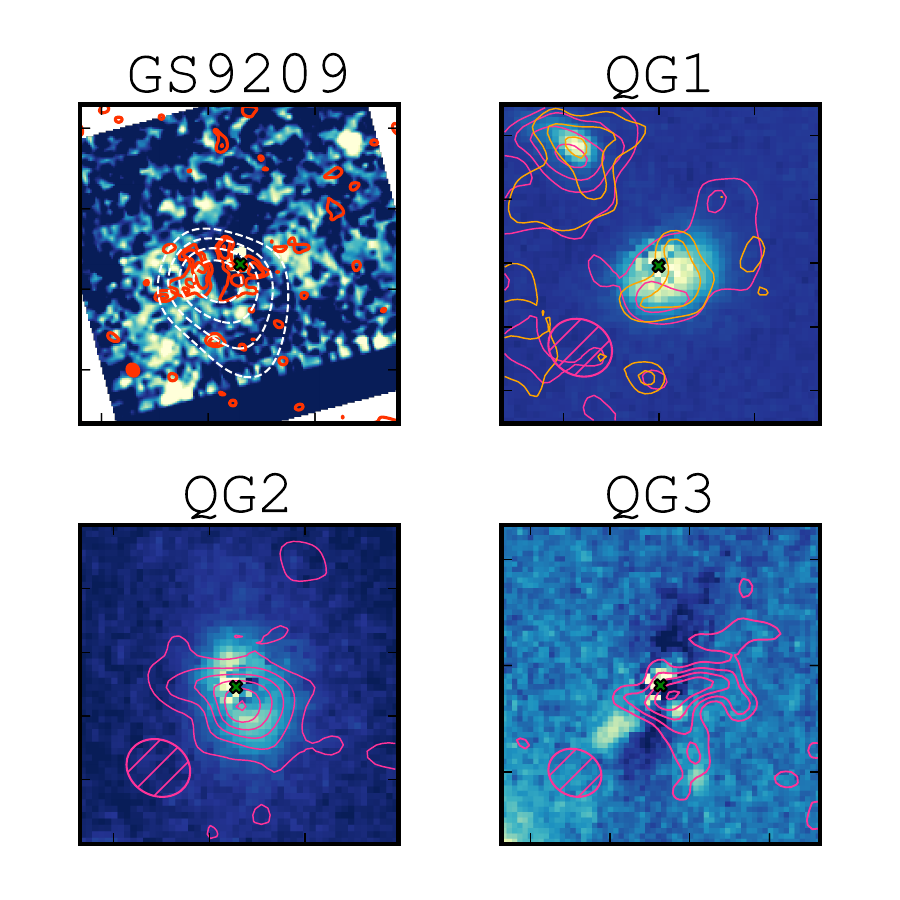}
    \caption{Background images: GALFIT residuals in NIRCAM/F356W for GS-9209 and F444W for QG1-2-3. [CII] and dust continuum contours are the same as in Fig.~\ref{fig:all}. Residuals for QG1 and QG2 were obtained with double Sérsic fitting and subtracting the main bulge component. White dashed contours for GS-9209 show the extent of the dust continuum detection after tapering. Dust continuum contours for GS-9209 at the native resolution are plotted in red for better contrast.} % \textbf{Residuals for QG1 and QG2 were obtained fitting two components, one for the primary and one for the secondary component, to avoid a hole at the galaxy core.}
    \label{fig:residuals}
\end{figure}

Local PSBs typically host strong tidal features and kinematics indicative of violent relaxation due to major, late stage mergers \citep{Zabludoff96, Yang04, Yang08}. While rare locally, they are expected to be more common at high redshift \citep{Tran04, Kriek16, Zahid16}.
Visual inspection of several of our four $z>3$ targets suggests disturbed morphologies or additional components beyond the main galaxy emission, possibly related to offsets in ALMA emission (continuum or \cii). We thus systematically inspected the morphology of our $z>3$ galaxies by fitting and subtracting a Sérsic component from the F444W image (we used F356W for GS-9209 due to the higher SNR) with \textbf{GALFIT}. Fitting the whole galaxies, including their extended halos, failed to converge, confirming structural complexity. We thus proceeded by constraining the best fitting Sérsic model only on galaxy cores, and subtracting it from the whole image with the same parameters. When the procedure revealed an additional component overlapping with the main Sérsic core (QG2 and QG1) we fit double Sérsic models to account for this. Strong residuals can be found in all cases, as shown in Fig.~\ref{fig:residuals}.

GS-9209 exhibits broadly extended and diffuse residual emission on the scale of 1.4$''$(11 pkpc; diameter), broadly consistent with the ALMA dust continuum counterpart.
For QG1 there is evidence of additional compact emission close to the core, in addition to the obvious QG1-Sat (see Figs.~\ref{fig:all} and ~\ref{fig:residuals}). The additional compact emission is best reproduced with a flat ($n=0.4$) source, lacking any substructure. QG2's residuals displays an additional component asymmetrically distributed around the main core. We fail to properly fit with any model, suggesting again diffuse stellar material. We estimate this extra component to contribute to 30-35\% of the total flux of QG2. For QG3 the residuals mainly show three distinct blobs, all to its southern side. We cannot confirm the redshifts of these QG3 blobs from [CII], but their projected proximity with the galaxy suggests association.

These structures seem to have interesting relations to the ALMA emission. In GS-9209, the ALMA continuum at 158$\mu$m rest is surprisingly extended and asymmetric, tracing the diffuse stellar residuals, indicating that part of the dust is in this diffuse component. 
In QG1 the [CII] and continuum signals that we find are quite remarkably overlapping with the stellar residual.
%While in QG1 we do not have secure detections of either [CII] or continuum at the expected JWST position, we see 3-4$\sigma$ peaks associated to the positive residuals  offset in the same direction from the core. Allowing for offsets extraction we find a corresponding continuum of $0.43\pm0.14$mJy (3$\sigma$) and a possible [CII] emission at the expected frequency with 4.1$\sigma$ significance, although this requires confirmation at higher SNR. 
For QG2 the ALMA [CII] detection appears to be offset from the main core along the major axis of the residual component.
%, although the offset is low in significance due to the large beam and modest SNR \textbf{and needs confirmation}. 
For QG3 the complex ALMA [CII] morphology suggests the presence of a tail or of an outflow, only partially overlapping with the stellar residuals.

Globally, we find striking evidence that all the $z>3$ quiescent galaxies in our sample exhibit complex stellar residuals as clear evidence of past mergers and interactions. This suggests that quiescent galaxies observed closer to or at their formation phase reveal complex stellar morphologies and that their residual ISM is similarly affected. 

\subsection{Evidence of dust heating not driven by star-formation}

In this section, we present evidence for enhanced dust heating beyond that expected from stellar radiation alone. Focusing on the two galaxies with robust $\rm{T_d}$ and $L_{\rm IR}$ constraints, we show that their bolometric dust luminosities cannot be explained by energy balance from dust-absorption stellar light, implying additional heating mechanisms. We explore possible scenarios and identify, in QG2, a candidate case of ISM heating driven by radio-mode feedback.

\begin{figure}[]
 \centering
    \includegraphics[width=\columnwidth]{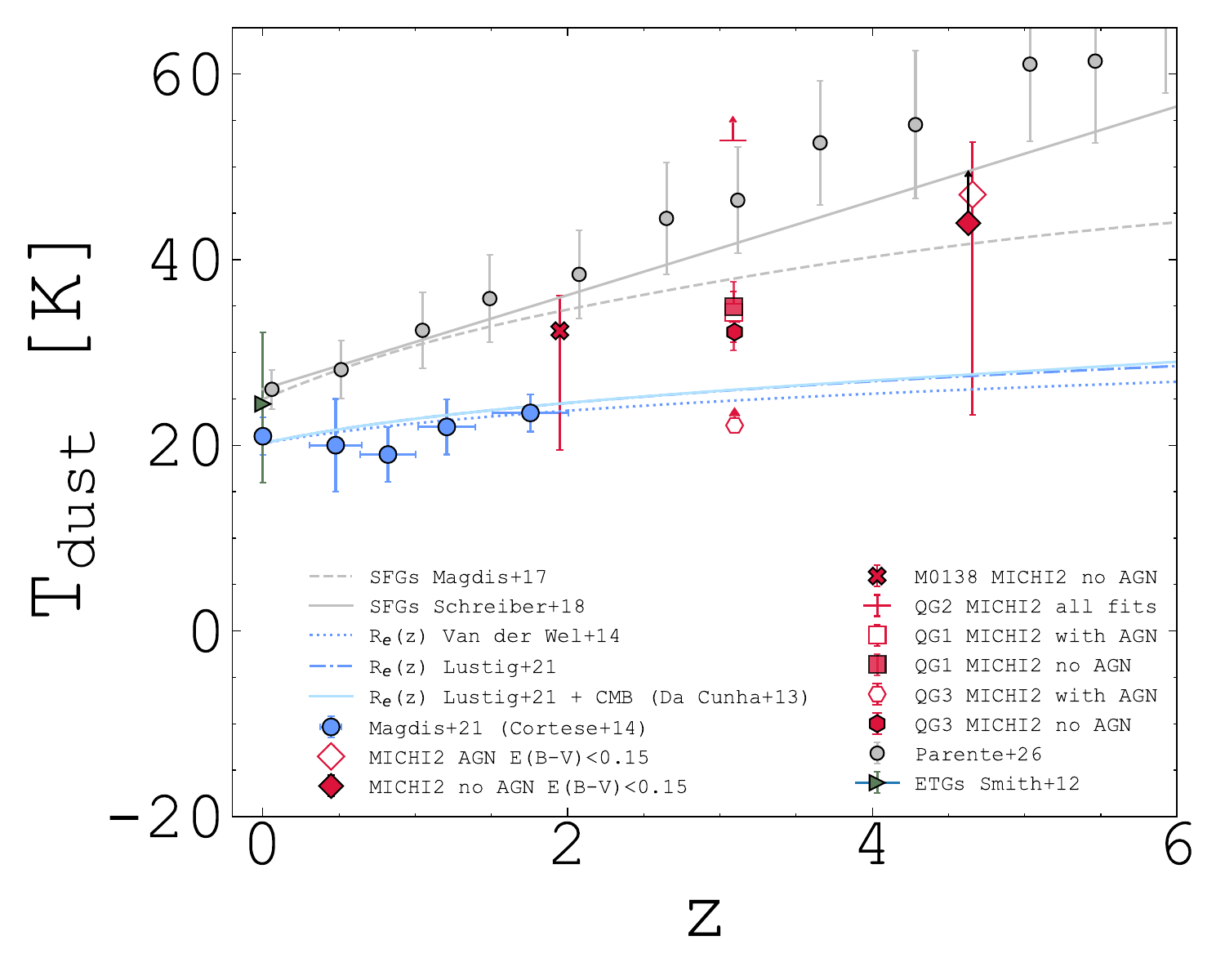}%
    % \caption{2 Figures side by side}%%
    \caption{Dust temperature as a function of redshift. Red points show our sample. Light blue circles show quiescent galaxy stacks from \cite{Magdis21}. The solid and dashed gray curves show the dust temperature evolution for star-forming galaxies from \cite{Schreiber18b} and \cite{Magdis17}, respectively. The light blue dotted and dot-dashed curves show the expected evolution of $\rm{T_{d}}$ from the size evolution of \cite{vdw14} and \cite{lustig}. The teal solid curve shows the trend expected from \cite{lustig} adding the effect of the CMB.}% \textbf{Dust temperatures from Xilouris et al 2004 are derived from black-body fits, assuming a dust emissivity of $\rm{\beta =1}$.}}
    \label{fig:tdust_evo}
\end{figure}

\cite{Ji2024} already discussed evidence for warm dust in GS-9209, based on the study of its MIRI emission and an ALMA dust continuum upper limit. We confirm this finding through the detection of dust continuum underlying the [CII] line, supporting a temperature of $\rm{T_{d}>44}$ K. This temperature is significantly higher than that expected from an intense interstellar radiation field produced in quiescent galaxies of increasing stellar mass surface densities due to their decreasing sizes at higher redshift, as can be seen in Fig.~\ref{fig:tdust_evo}, following the modeling of \cite{Chanial2007} in \cite{Magdis21}.
We also find evidence for even warmer dust in QG2 (see Sect.~\ref{sec:sedfitting_qg2}), based on its bright MIRI emission, an ALMA upper limit and its radio detections. Its $\rm{T_{d}}$ lower limit exceeds the that of coeval star-forming galaxies. To a lesser extent, QG1 also shows a warmer dust compared to expectations for quiescent galaxies at its redshift.
 
These results are somewhat unexpected. Stacking analyses of lower-redshift quiescent galaxies have yielded very cold dust temperatures \citep[e.g.,][]{Gobat18, Magdis21}, consistent with heating from residual star formation or evolved stellar populations. In such systems, the presence of massive spheroidal components is associated with low star formation efficiencies, and $\rm{T_{d}}$ —being sensitive to the ratio $L_{\rm IR}/M_{\rm d}$— remains correspondingly low. On the other hand, low-redshift PSB galaxies occasionally exhibit warm dust, indicating that elevated dust temperatures are not unprecedented in recently quenched systems \citep{Smercina2018}. \cite{Ji2024} attribute the warm dust in GS-9209 to a recent, spatially extended episode of star formation.
Motivated by this, in what follows we test whether reprocessed stellar radiation can explain the observed dust heating and find it insufficient to reproduce the inferred IR luminosities. We then compare our high-redshift galaxies to local post-starburst (PSB) systems and explore alternative heating mechanisms.

\subsubsection{Excess LIR compared to reprocessed light}
\label{sec:cirrus}

We compute the UV–optical light that should be re-emitted into the FIR following this procedure: we interpolate the optical SED of our galaxies, we de-redden it assuming the published value of the optical extinction (Av) and we compute the absorbed luminosity by integrating the difference between the intrinsic and the reddened SEDs over 912 \AA\ to 2 $\mu$m rest.
For GS-9209 we use the optical SED integrated over the whole galaxy \cite{Ji2024} (see their Fig.~7), consistent with the best fit parameters of \cite{Carnall23}, and $\rm{Av=0.04^{+0.05}_{-0.03}\, mag}$ from \cite{Carnall23}. For QG1 and QG2 we take their optical SEDs and Av values from \cite{Kubo21,Kubo22}.
As can be seen in Fig.~\ref{fig:Lcirrus_vs_LIR} (left panel), for QG2 and GS-9209 the measured LIR exceeds the reprocessed stellar light by an order of magnitude or more. For GS-9209, depending on whether or not we include an AGN torus in our FIR SED fitting, the excess reaches nearly 2 dex.

A  visual exemplification of this phenomenon comes from the SSA22 field, where QG2 has similar F444W, F2100W and $L_{\rm IR}$ luminosity to ADF22.A4, a highly star-forming galaxy in the proto-cluster. Still their near-IR colors (F200W-F444W) are strikingly different by 2.4 magnitudes ($\times10$), QG2 being much bluer and consistent with negligible attenuation.

%\textbf{We here show that such a large LIR for QG2 does not come from obscured star formation by comparing its observed F200W-F444W color to other dusty star-forming galaxies in the same environment. We take galaxy ADF22.A4 as a reference as it stands out in terms of similarities with, since it is detected in X-rays, it is bright in JWST/MIRI F2100W, )(contrary to other galaxies in the same field) and it has ADF22.A4 has a $\rm{log(LIR/\Lsun)= 12.3^{+0.2}_{-0.1}}$, only a factor of 2 larger than what we find in QG2. Its SFR is $\rm{SFR_{IR}= 380^{+290}_{-70}\, \Msun yr^{-1}}$ \citep{Umehata17}. Recently, \cite{Umehata26} measured F200W-F444W colors for the dusty star-forming galaxies with the brightest and dustiest cores in the ADF22 group. ADF22.A4 has $\rm{F200W-F444W\sim 3 \,AB\,mag}$, computed by performing aperture photometry on its core using an aperture diameter of d=0.15". For QG2 we obtain $\rm{F200W-F444W = 0.66 \pm 0.07 \,AB\,mag}$ using a Kron aperture. Uncertainties already account for the lack of PSF-matching between the F200W and F444W filters in our measurement. Thus QG2 has a 2.3 mag smaller JWST color against an LIR that is only a factor of 2 lower than that of ADF22.A4. We conclude that QG2 does not display sufficient reddening to explain its large LIR with on-going dust-obscured star formation.}

\subsubsection{Decoupling LIR from star formation}

\begin{figure*}[]
 \centering
    \includegraphics[width=\textwidth]{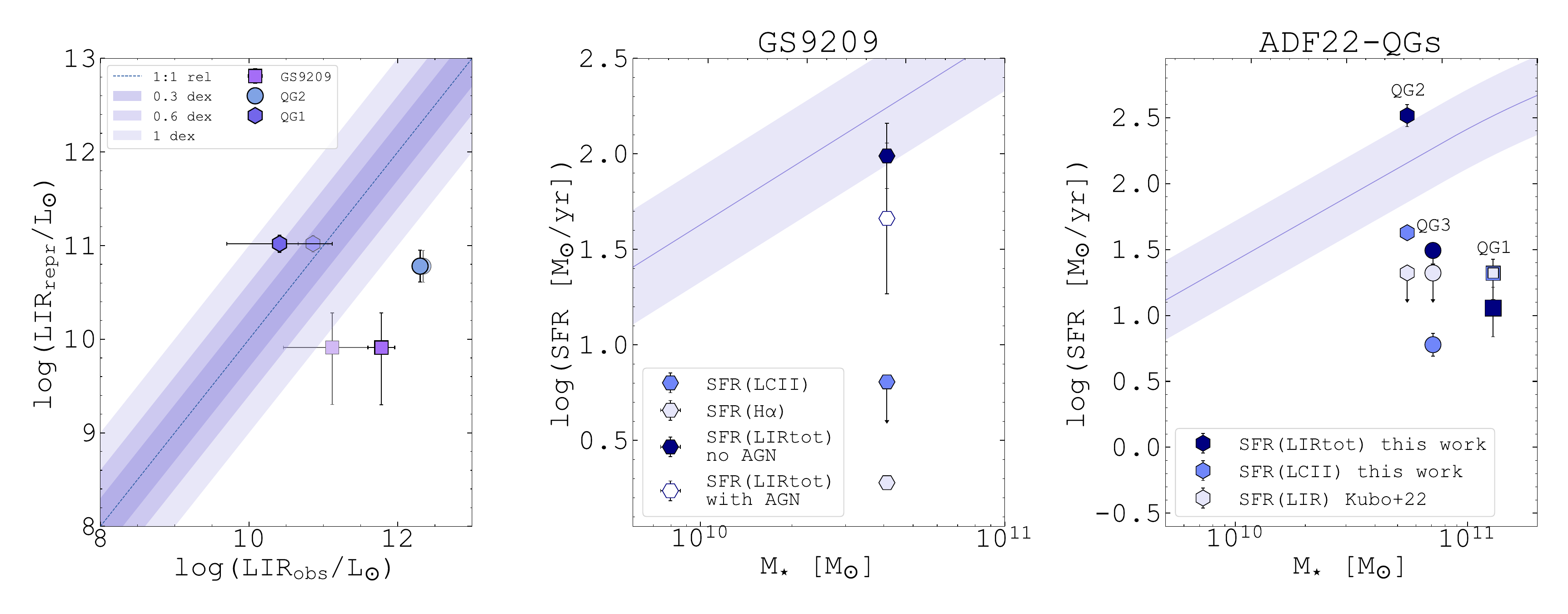}%
    % \caption{2 Figures side by side}%%
    \caption{\emph{Left}: Comparison between the IR luminosity expected from reprocessed light from the UV and the observed IR luminosity. \emph{Middle}: SFR of GS-9209 with respect to its coeval MS \citep{Schreiber2015} according to different SFR constraints. \emph{Right}: same as middle panel but for the quiescent galaxies in the AzTEC14 group.} 
    \label{fig:Lcirrus_vs_LIR}
\end{figure*}

Fig.~\ref{fig:Lcirrus_vs_LIR} compares multiple SFR diagnostics for our galaxies in Main-Sequence (MS) diagrams at $z=4.7$ (middle panel) and for the $z=3.09$ (right panel). Remarkably, the IR luminosities of GS-9209 and QG2 are high enough to locate them within the MS, while their [CII]-based SFRs fall below it. The discrepancy is even more pronounced for GS-9209 when adopting the H$\alpha$-based SFR, which lies $\sim$1.5 dex below the MS and may represent an upper limit due to AGN contamination \citep{Carnall23}.
This further supports that $L_{\rm IR}$ is not a reliable SFR tracer in these systems, consistent with \cite{Smercina2018, Wild25}. Moreover, being the dust-reprocessing of stellar radiation insufficient, it shows that additional boosting of the IR luminosity is required. Regular JWST colors for their main stellar bodies further rules out heavily obscured star-forming components. Finally, we note that the [CII]-based SFR of QG2 lies $\sim$4 times below the MS, placing it in the Green Valley. In the following sections, we provide elements to suspect that its \Lcii might be boosted by processes other than star formation. 

\subsection{Comparing $z>3$ quiescent galaxies to local post-starbursts}

Given that MQGs at high redshift are observed shortly after quenching, it is instructive to compare them with lower-redshift PSB systems.  We adopt as a reference \cite{Smercina2018} (hereafter S18) who studied 33 local E+A PSB galaxies combining infrared photometry, Spitzer IRS spectroscopy, [CII] and [OI] emission lines from Herschel PACS. These observations enabled constraints on total gas and dust masses, \Lcii and $T_{d}$, and thus they are particularly suited for our study. S18 reports high central surface densities and even more compact IR cores of warm dust, heated by a high-intensity soft radiation field from A-type stars. Such a radiation field boosts their \LIR driving strong [CII] deficits. $H_{2}$ spectroscopy further indicates widespread shock-dominated regions suggesting, overall, that their ISM is supported against collapse by a continued injection of turbulent or mechanical heating. 
Here we list their main findings and highlight key differences with our sources.

\begin{figure}[]
 \centering
    \includegraphics[width=0.45\textwidth]{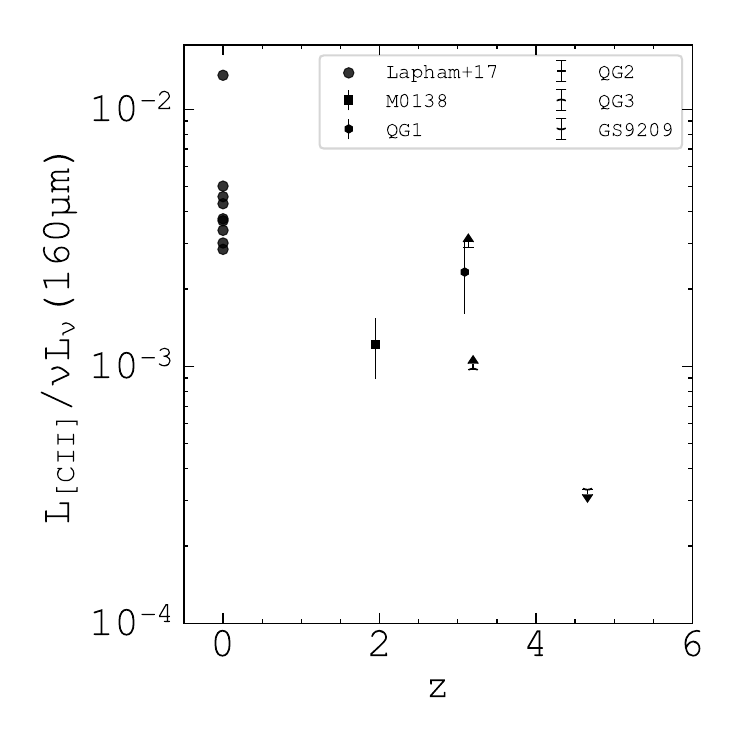}
    \caption{
    The ratio between the [CII] luminosity and its underlying continuum. Local QGs from \cite{Lapham2017} are shown alongside our targets.}
    \label{fig:lcpl160}
\end{figure}

\begin{itemize}
    \item[1)] \textbf{Dust is warm:} on average, SEDs in S18 peak at $\rm{\lambda_{peak}\approx 70-75\, \mu m}$ instead of $\sim$100 $\mu m$ of normal star-forming galaxies, implying warmer grains and a higher intensity radiation field, comparable to (U)LIRGs, corresponding roughly to $T_{\rm d}\sim30$--35~K and $<U>\sim30$--40, respectively. This heating is attributed to dense stellar regions dominated by A-type stars where a "high-soft" radiation field is intense enough to heat dust grains without a significant FUV contribution from on-going star formation. In contrast, we find $\rm{\lambda_{peak}\approx 40-48\, \mu m}$ and thus even warmer dust temperatures. Although the proximity of our galaxies to their quenching epoch suggests a potentially stronger soft-UV field from A-type stars, in QG2 and GS-9209 the \LIR expected from reprocessed stellar light falls far short of the observed one. Additional mechanisms that could provide such heating are widespread shocks, heating at formation from AGB stellar envelopes or AGN radio-mode feedback, which we explore in the following sections; \\

    \item[2)] \textbf{The ISM in S18 is mainly distributed in compact IR cores:} typically on scales smaller than 1/3 of the optical size, approaching (U)LIRGs-like sizes at rest 8 $\mu m$ \citep[$\sim500$~pc,][]{Tacconi06} suggestive of dense, compact star-forming progenitors.
    In our case, M0138 ($z=2$) shows a compact dust core in the bulge but our data only provide a $2\sigma$ upper limit of 0.35$''$ (3 kpc) on its FIR size. Optical source-plane morphology reveals a regular disk with no clear signs of recent interactions \citep{Newman2025}. For the rest of our sample, dust continuum and/or [CII] emission is extended, on scales comparable to or larger than the optical size, at odds with S18.\\ 
    While interpreting the bulk of \Lcii\ as purely tracing molecular gas is uncertain due to possible contributions from the ionized phase, the extent of our detections suggests that we might be witnessing the redistribution of gas following the major assembly event of our galaxies, rather than the formation of very compact IR cores as in local PSBs;\\
    
    \item[3)] \textbf{[CII] deficits:} As can be seen in \cite{Zanella18}, \Lcii/LIR tend to drop significantly below 0.1\% for galaxies with $LIR>10^{12}\Lsun$, thus exhibiting a [CII] deficit compared to their high LIR. [CII]/LIR ratios in S18 show a rather large variation\footnote{We use TIR/\LIR = 1.06 to convert from their 3–1100 $\mu m$ TIR to our 8–1000 $\mu m$ \LIR, as in S18.}, typically around 0.1--0.2\%, with a few cases reaching down to 0.01\% and many upper limits indicating that up to half of their sample could be compatible with ULIRG-like [CII] deficits. In our sample, explicit calculations are feasible only for QG2 and GS-9209 (and to a lesser extent for QG3) yielding [CII] deficits $\sim$0.01--0.05\%, characteristic of ULIRGs.
    This again suggests that quiescent galaxies closer to their formation epoch share more properties with local PSB systems than they do with fully mature ETGs.
    This is further supported by Fig.~\ref{fig:lcpl160}, where $\Lcii/L_{160\mu m}$ --a less model-dependent proxy for \Lcii/LIR-- appears to decline compared to local gas-rich ETGs for which such measurements are available in \cite{Lapham2017} and \cite{Temi2009b}. This behavior is not seen in the LCO(1-0)/L[CII] ratio (Fig.~\ref{fig:Lcii_Lco}). We find no evidence for evolution in the [CII]-to-dust ratio itself. Instead, our data point towards increased [CII] deficits in QGs at high-z, where the rising dust temperatures and rising IR luminosities are not matched by a corresponding rise in [CII] luminosity. We note here that we have only 5 galaxies at our disposal, therefore we cannot properly assess the scatter in [CII] deficit and we do not know to which extent such galaxies are representative of the whole population of high-z quiescent galaxies (for example QG2 appears to be a rather peculiar system). Larger samples are needed to populate the high-z regime of this diagram.
    Since these galaxies are much closer to their main formation epoch, we tend to interpret the trend in [CII] deficit with redshift shown in Fig.~\ref{fig:lcpl160} as a sign that high-z MQGs have not yet fully relaxed into quenched ellipticals and that their ISM conditions are more similar to those of local PSB, if not more extreme;\\

    \item[4)] \textbf{ \LIR not linked to SFRs:} 
    S18 argue that steep stellar surface brightness profiles lead to overestimated SFRs when inferred from $L_{\rm IR}$ and [CII]. The SFR in their galaxies can decline by $>2$ dex since the starburst while UV emission from post-burst populations can sustain elevated IR luminosities for up to $\sim$1.5 Gyr without ongoing star formation \citep[see also][]{Wild25}. As discussed in Sect.~\ref{sec:cirrus}, we find that $L_{\rm IR}$ in our galaxies exceeds energy-balance expectations by $>1$ dex, where $L_{\rm IR}$ is neither consistent with residual star-formation nor with attenuated light from older stars.
    %assuming literature $\rm{A_V}$ estimates are reliable. 
    We further note that our inferred $L_{\rm IR}$ exceed by 1-2 dex typical values reported for local PSBs in S18 and \cite{Wild25}. We therefore tend to exclude stellar radiative heating as the primary driver of the inferred $L_{\rm IR}$ in favor of alternative mechanisms as outlined above;\\ 
    
    \item[5)] \textbf{Significant dust reservoirs and low SFE:} dust masses in S18 are as high as $5 \times 10^{8} \Msun$ corresponding to a broad range of \dgr\ ($25\leq \dgr \leq 400$) when combined with CO-based gas masses. They argue for their galaxies being dominated by molecular gas after accumulating significant dust destruction. Despite not having direct constraints on \dgr\ at z$>$2, we obtain relatively low molecular gas fractions, ranging from 0.2\% to 5\%, with the exception of \fg$\sim$25\% for QG2 which we believe is undergoing additional [CII] boosting as discussed in Sect.~\ref{sec:qg2_boosting};\\

    \item[6)] \textbf{Presence of turbulence and shocks in the ISM:} S18 find evidence for warm gas ($T>50$ K) from elevated H$_2$/7.7,$\mu$m ratios, indicative of a turbulent ISM heated by low-velocity magnetohydrodynamic shocks. In 95\% of sources with MIR spectroscopy, H$_2$ emission is more extended than in star-forming galaxies, with $\rm{L{H_2}/TIR}$ consistent with shock-dominated regions. PAH ratios show little evidence for small-grain destruction, favoring shock- or turbulence-driven heating of the molecular gas component. Additionally, several sources exhibit strong [OI] 63 $\mu$m emission relative to [CII], as predicted by shock models. Two potential drivers of turbulent heating include recent galaxy interactions/mergers and radio-mode AGN feedback via jets. Although increased ISM turbulence is very much plausible for our galaxies, detailed MIRI spectroscopy for our sources is currently lacking.\\

\end{itemize} 

\subsubsection{Plausible dust heating mechanisms}

In this and the following subsection, we speculate on what processes could account for dust heating in excess of stellar radiation. While a definitive identification might be difficult and beyond the scope of this work, it is interesting to list several mechanisms that can play a role in our galaxies:

\begin{itemize}
    \item[A)]  \textbf{Shocks:} shocks induced by tidal interactions or mergers —widespread in our sample— can generate turbulence and heat dust, as observed in local PSBs and AGN hosts \citep{Haidar2024,Haidar2026}, without significant dust destruction. Notably, QG2 and GS-9209, which exhibit the highest $T_{\rm d}$ and strongest [CII] deficits, also show disturbed stellar morphologies, including lopsided structures and extended stellar halos on $\sim$10–15 kpc scales. Furthermore, at high redshift galaxies are denser and more compact so, possibly, shocks might be even more efficient than locally;\\ 
    
    \item[B)] \textbf{AGB stars:} 
    with stellar ages of $\sim$200–800 Myr, our galaxies are likely near the peak of AGB dust production. Dust in AGB envelopes can reach temperatures $>400$ K \citep{Lagadec2005} and its contribution to the emission from the stellar surface is ingrained in stellar templates, thus difficult to disentangle with simple energy balance considerations;\\

    \item[C)]  \textbf{Collisional heating by hot plasma:} dust heating via collisions with electrons in a hot plasma phase is another possibility \citep{Bocchio2013,Drozdov2019}. However, there is currently no X-ray evidence for a hot intracluster medium in the SSA22 protocluster \citep{Lehmer2009}, and dust in local ETGs remains typically cold despite being embedded in a hot, diffuse X-ray halo;\\
    
    \item[D)] \textbf{Different dust properties:} it is also possible that an altered grain size distribution and/or emissivity could potentially explain our data. Additional photometry in the MIR and (sub)mm regimes could further clarify this point.

\end{itemize}

\subsubsection{QG2: Candidate test-case of on-going radio-mode feedback at z=3}
\label{sec:qg2_boosting}

\begin{figure}[]
 \centering
    \includegraphics[width=0.8\columnwidth]{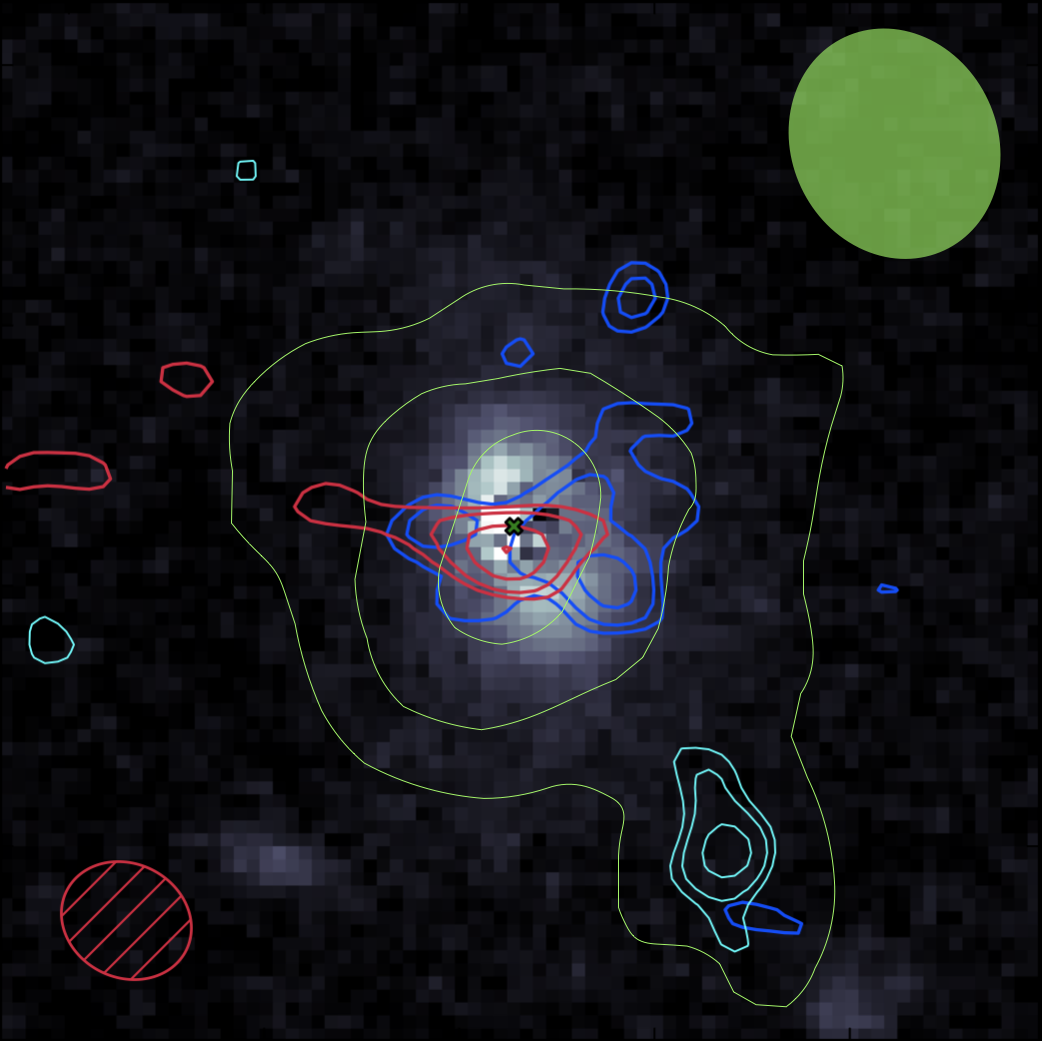}%
    \caption{Background image: F444W residuals. [CII] (red and blue) and dust continuum (cyan) contours are plotted at 2.5, 3 and 4 $\sigma$ for clarity. Green contours show the 6 GHz radio emission at 3,6,9 $\sigma$ from \cite{Kubo22}. The major axis of the stellar lopsided component appears to be aligned with the radio lobe and the dust continuum contours. The red and blue components of [CII] are obtained by integrating the signal over the wings of the [CII] spectrum. } 
    \label{fig:QG2_jet_cii_rotation}
\end{figure}

For QG2, SED fitting of the 21 $\mu$m MIRI detection combined with ALMA upper limits yields a robust lower limit of $T_{\rm d}>52$ K, exceeding typical values for star-forming galaxies at similar redshifts \citep{bethermin15,Schreiber2015}. MICHI2 fits rule out a dominant AGN torus component, and the spatially resolved 21\,$\mu$m emission indicates that the warm dust is distributed on galactic scales, rather than in an unresolved nuclear component. F444W residuals reveal a lopsided morphology consistent with a past major merger. The [CII] line profile in Fig.~\ref{fig:QG2_jet_cii_rotation} (blue and red contours) suggests rotation or two merging components, although the low SNR prevents a clear assessment. 6 GHz radio contours at $3\sigma$ in \cite{Kubo22} reveal an elongated and diffuse component which we detect in continuum at 158\,$\mu$m rest at $4\sigma$, without any stellar counterpart in JWST images up to $\sim7\,\mu$m. The radio power measured on the core of QG2, AGN-like optical line ratios, and the morphology aligned with the radio axis suggest the presence of a putative radio hot-spot. The lack of NIRCam and MIRI counterparts at its position disfavors jet-induced dust-obscured star formation. The absence of a 1.1 mm counterpart for our 158 $\mu$m rest detection, however, disfavors synchrotron emission and suggests potential jet-induced dust heating.
Thus, the high $L_{\rm [CII]}$ measured on QG2, compared to weak 1.1mm dust continuum and CO(3-2) emission, suggests [CII] powering by shocks or turbulence driven by a merger and/or a radio jet. We cannot exclude, however, a contribution to \Lcii from the [OIII] ionized outflow observed in the optical. 

In summary, QG2 exhibits: (i) warm dust ($T_{\rm d}>52$ K); (ii) evidence for on-going SMBH accretion from its observed and bolometric X-ray luminosities ($\rm{log(L_{2-10\,Kev}/erg\,s^{-1})=43.2\pm0.1}$ and $\rm{log(L_{bol}/erg\,s^{-1})\approx46.0}$, respectively) and a fairly high Eddington ratio \citep[$L_{\rm bol}/L_{\rm Edd}\sim0.4$,][]{Kubo22}; (iii) AGN-driven photoionization and ionized outflows; (iv) disturbed stellar morphology consistent with a major merger and (v) a potential radio jet.
The X-ray detection suggests that the mechanism producing a collimated radio jet may still be in place and that could provide enough energy to radiatively heat the ISM of the galaxy. Higher SNR observations are required to establish whether the [CII] emitting gas in QG2 is in regular rotation.

Since the dynamical timescales typically involved in a major merger before complete coalescence are longer than the typical AGN detectability in the X-rays (10$^{8}$ yr vs. 10$^{5-7}$ yr, respectively), we interpret these properties as the outcome of a merger-triggered accretion episode that both fueled the AGN and heated the residual ISM through a combination of shocks, turbulence, and jet-driven energy injection. While the pinpointing the main driver of the dust heating will require additional observations, these results support a scenario in which major mergers play a central role in high-redshift quenching, driving rapid gas depletion, ISM heating, and episodic AGN feedback.

% \begin{figure*}[]
%  \centering
%     \includegraphics[width=\textwidth]{fmol_Chabrier_updated_10Apr26_Cii.png}%
%     \caption{Molecular gas fraction evolution of quiescent galaxies with redshift. \textbf{Yellow markers highlight the trends resulting from dust continuum studies.} } 
%     \label{fig:fgas_vs_z}
% \end{figure*}

\subsection{A point on the evolution of the gas content in quiescent galaxies at $z>3$}

\begin{figure*}[]
 \centering
    \includegraphics[width=\textwidth]{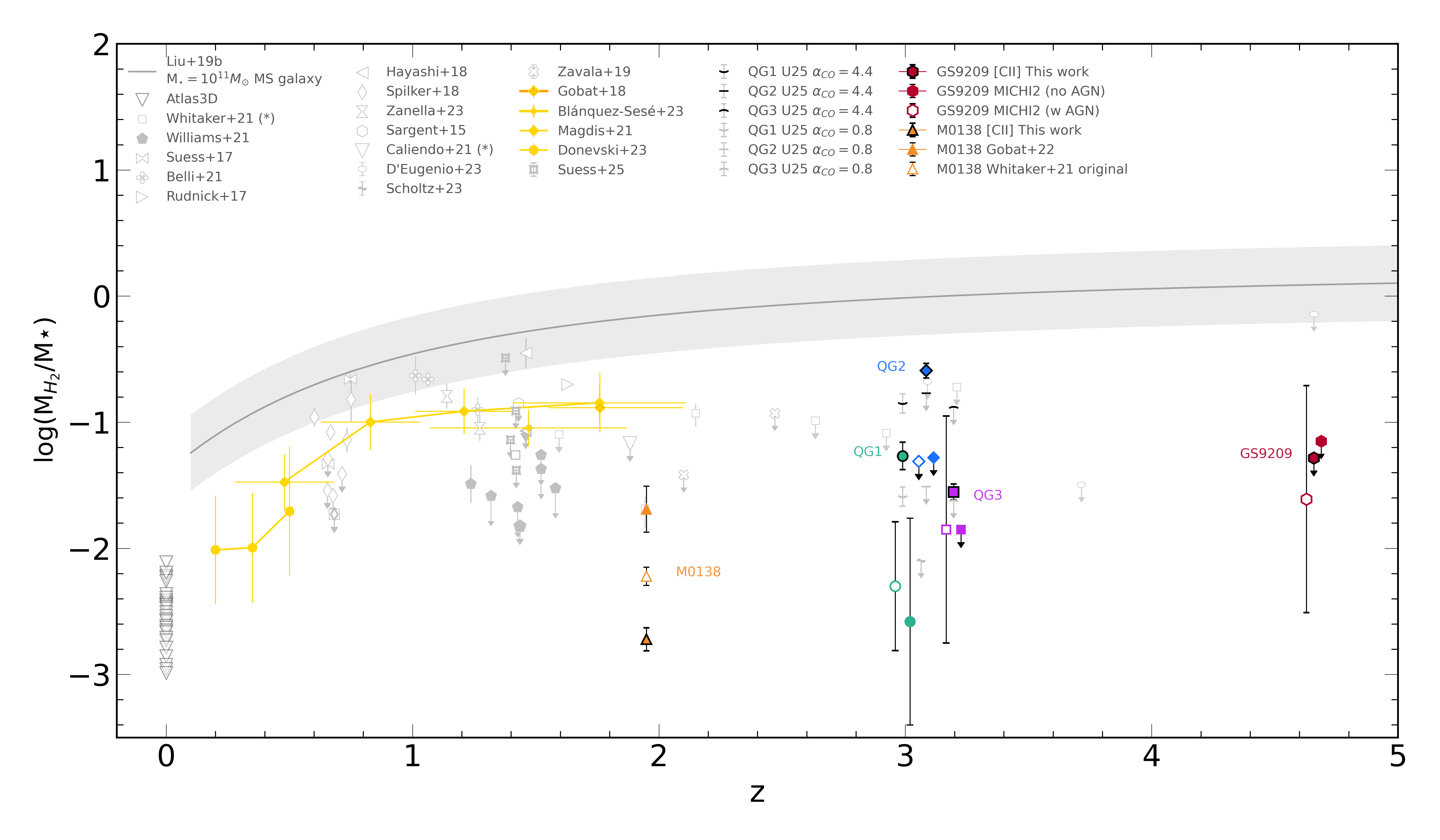}%
    % \caption{2 Figures side by side}%%
    \caption{Molecular gas fraction evolution of quiescent galaxies with redshift. For each of our $z>3$ targets we show gas fractions obtained based on \Lcii (black-outlined filled symbols), FIR SED fits including and excluding the AGN component (empty and filled symbols, respectively). Markers referring to same target, or to targets at a similar redshift, have been slightly shifted in redshift for clarity. Yellow markers highlight the trends resulting from dust continuum studies. The gray solid line shows the gas fraction evolution for a main sequence galaxy with $M_{\star}=10^{11} \, \Msun$ from \cite{Liu19} with a 0.3 dex scatter (gray shaded area).} 
    \label{fig:fgas_vs_z}
\end{figure*}

Fig.~\ref{fig:fgas_vs_z} compares the gas fraction estimates in our targets with literature values, showing our results from both [CII] and dust. 
Most of our galaxies show low gas fractions ($\lesssim$ a few percent), below the average trends reported by \cite{Magdis21} and \cite{Gobat18}, and consistent with ALMA stacking results from \cite{Adscheid2025}. The exception is QG2, whose [CII]-based \fg reaches $\sim$25\%, although this is not supported by dust measurements and may reflect [CII] enhancement from shocks or AGN-driven heating, as discussed in the previous section.
We also note that the potential presence of warm dust in these systems complicates the use of fixed dust temperatures to infer dust and gas masses. Accounting for large temperature uncertainties can significantly broaden the inferred gas fraction range.

\subsection{The origin of [CII] emission in high-z QGs}

It is not straightforward to assign the origin of [CII] to a given phase of the ISM of our galaxies. Assessing the contribution of ionized gas to \Lcii is particularly relevant to understand the plausibility of our assumption on \alphacii and to understand which ISM phase are we considering when comparing our \cii maps to the emission from other galaxy components. Given the lower ionization potential of carbon (11.3 eV) compared to hydrogen (13.6 eV), [CII] is produced both in the ionized and the neutral gas components. Most studies have concluded that the bulk of [CII] emission originates in neutral regions, both in the local universe and at high-z, with a contribution from ionized gas typically ranging around $\sim$5-30\% \citep{Wolfire22}.
In star-forming galaxies, [CII] primarily arises from classical photodissociation regions (PDRs), where FUV radiation from young O and B stars heats gas and dust via the photoelectric effect.\footnote{PDRs can also be produced in planetary nebulae or reflection nebulae, as well as protostellar and protoplanetary disks but the limited spatial extent and luminosities of these components make them negligible at the redshifts considered here.} In classic PDRs, the gas and dust are heated by photoelectric effect, namely by the ejection of electrons from small dust grains and PAHs, which provides partners for collisional excitation of fine-structure line emission (free electrons) and continuum emission (dust grains). 
[CII] can also originate in X-ray-dominated regions (XDRs), where X-rays from an AGN accretion disk or from X-ray binaries enhance gas heating and cooling efficiencies. However, this enhancement is limited to the immediate surroundings of the heating source, with typical sizes of a few hundreds of parsecs \citep{Wolfire22}. In our sample, three of the four [CII]-detected galaxies (M0138, QG1, QG3) are undetected in X-rays, while QG2 hosts a luminous AGN but shows extended [CII] emission on $\sim$3 kpc scales. Additionally, QG1 and QG3 show an extended [CII] tail far outside the stellar body of the galaxy, likely due to an outflow or a tidal event. Although low-level AGN activity cannot be ruled out in all cases, the available evidence suggests that XDRs do not dominate the observed [CII] emission.\\ %\textbf{Comparing \Lcii to the luminosity of [OI] 63$\mu$m, a the major coolant in XDRs, would provide further insights.}
%[CII] can also originate from ionized gas. %Ideally, the ionized gas contribution to \Lcii could be constrained via the [CII]158/[NII]205 ratio \citep{Schreiber22}. \cite{Lapham2017} find elevated [NII]122/[CII] ratios in about 20-22\% of their sample, with a maximum of $\sim$50\% if a threshold of [NII]122/[CII]$\gtrsim$0.12 from \cite{Malhotra2001} is adopted. PSBs in S18 show a decreasing [CII]/PAH ratio as dust grains become more ionized, indicating a larger contribution from ionized gas compared to local star-forming galaxies.
Ideally, the ionized gas contribution to \Lcii could be constrained via the [CII]158/[NII]205 ratio \citep{Schreiber22} which, however, is unavailable for our sample. 
We can, however, take a look at local ETGs: \cite{Lapham2017} find low [CII]/[NII] ratios in about 25\% of their sample indicating enhanced gas ionization preferentially in systems with larger dust temperatures. Our results for QG2 suggest that a significant part of its [CII] luminosity could indeed arise from ionized gas, potentially linked to the same process leading to its high dust temperature. %PSBs in S18 show a decreasing [CII]/PAH ratio as a function of 7.7$\mu$m/11.3$\mu$m PAH ratio, consistent with a strong reduction in photoelectric efficiency as grains become more ionized.
Dissipation of mechanical energy (turbulence and shocks) in highly turbulent conditions --such as galaxy mergers or AGN-driven outflows-- was also proposed as a [CII] powering source \citep{Appleton13, Appleton17, DeBreuck22}. While S18 argue for low-velocity shocks as a viable source of heating in local PSBs, QG2 also hosts an [OIII] outflow. Thus, comparing the spatial distribution of the [OIII] emission line relative to [CII] will be useful to assess any role of the outflow in powering the [CII] line. Finally, measuring [CII]/PAH ratios will be useful to assess any excess [CII] emission relative to what expected from classical PDRs.% \textcolor{red}{make coherent with S18}

\section{Summary and Conclusions}\label{sec:summary}

We present ALMA observations of the [CII] line emission and its underlying continuum in 5 massive quiescent galaxies at $2<z<4.7$. Lacking a consensus on the actual ISM content in this class of galaxies, we exploit the first harvesting of deep high-z [CII] observations to probe their gas reservoirs with unprecedented sensitivity to study their ISM properties and cosmic evolution.
Complementary archival JWST NIRCam imaging, and MIRI where available, enables a detailed analysis of the host morphologies and SED modeling from the near- to mid-IR, providing additional constraints on their dust content. We summarize our results as follows:

\begin{itemize}
    \item[i)] We obtain the first [CII] detections in QGs in the distant Universe: three secure (M0138, QG2, QG3), one tentative (QG1), and a deep upper limit for GS-9209. Dust continuum is detected in GS-9209 and M0138, and tentatively in QG1;
    
    \item[ii)] We attempt a first cross-calibration of \alphacii for QGs in the distant Universe, finding no evidence for a strong evolution. For QG2 we measure the CO/[CII] ratio finding it consistent with both local QGs and star-forming galaxies. This provides reliability to attempts at estimating the gas content of distant QGs using local conversion factors, modulo the assumption of \dgr\ and \alphaco. We note, however, that we cannot rule out a mild increase in \alphacii of a factor of 2--3 in galaxies with a suppressed SSFR. Although this does not represent a strong evolution, it could significantly impact integration times for faint sources and should be taken into account when planning future observations; 
    
    \item[iii)]
    Our data suggest a downward evolution of \Lcii compared to the IR continuum, mainly linked to the pronounced [CII] deficits in high-z QGs;
    
    \item[iv)] All $z>3$ QGs in our sample show widespread merger signatures revealed after subtracting their main Sérsic bodies, hinting at complex and chaotic early times following their formation episodes;
    
    \item[v)] We find further evidence for warm dust \citep[supporting][]{Ji2024} together with evidence of dust heating not due to stellar radiation absorbed by dust. The IR of QG2 and GS-9209 violate the expectations from reprocessed radiation from dust by over an order of magnitude;
    
    \item[v)] The measured LIR are so large that these galaxies would be placed in the Main Sequence, despite convincing evidence for their quiescence from literature studies and JWST colors. This implies that IR luminosities cannot be used to estimate SFRs for these sources, as advocated in lower redshift studies;
    
    \item[vi)] Plausible heating mechanisms include merger-driven shocks and production in AGB stellar envelopes, consistent with their complex morphologies and their young stellar ages ($<1$ Gyr) and altered grain properties; 
    
    \item[vii)] 
    QG2 likely requires further explanation as it shows excess gas mass from [CII] relative to its dust mass, very warm dust temperatures, and potentially a very high gas-to-dust fraction. If the presence of a radio jet is confirmed, we suggest that there could be radio-mode feedback enhancing \Lcii and dust heating in this galaxy, representing the first known case to our knowledge;
    
    \item[viii)] A careful comparison with local PSBs reveals several similarities (warm dust, evidence for mergers, heating and turbulence injection by shocks) but also key differences, including potentially stronger [CII] deficits, more extended gas and dust reservoirs, and non stellar dust heating; 
    
    \item[ix)] We discuss the origin of [CII] emission, considering contributions from both PDRs and XDRs, and from molecular, neutral and ionized gas phases.
    \end{itemize}

This work highlights a number of unexpected findings that, at the same time, raise key questions on how these could be confirmed and best understood. In particular, the possibility of non-stellar ISM heating requires further confirmation. One possible way to distinguish between shock-dominated regions and AGB stars is assessing PAH line ratios and $\rm{H_{2}}$ rotational emission via MIRI spectroscopy and to verify the presence of AGB features in the rest-frame NIR, ideally with much larger samples. Furthermore, improved constraints on the FIR SED peak are essential to make these results less model dependent. Lastly, significantly larger samples are needed to determine whether these findings are representative or whether they pertain to individual outliers.

We identify a potential case of direct ISM heating from radio-mode feedback, as predicted by models \citep{Ciotti2007,Ciotti2010}. A systematic study of high-$z$ radio AGN in quiescent hosts could test this scenario by probing correlations between elevated dust temperatures and enhanced [CII] luminosities. 

More broadly, while being overall gas-poor, we highlight substantial variations in ISM content of our sample. As such, we speculate that different processes might be at play in different galaxies, and/or that there could be strong fluctuations in the ISM content. Much larger, homogeneous samples are needed to assess the scatter in gas and dust masses of high-z QGs.
%Because of these phenomena and the likely presence of substantial scatter, it might be premature and difficult to draw a clear-cut picture of the average ISM content of high-z QGs. 

Another notable result is the pervasive evidence for mergers, revealed by significant positive residuals after subtraction of the main Sérsic profiles. It remains unclear whether this high incidence reflects the unprecedented sensitivity and spatial resolution of JWST, or instead if it represents an intrinsic property of high-z QGs. Likewise, the potential dependence on stellar mass or environment has yet to be established. These questions are now directly addressable through a systematic exploration of existing JWST archival data (Bruckmann et al., in preparation).
    
\section{Acknowledgments}
We are grateful to Mariska Kriek for suggesting to look at AGB stars as a source of dust heating and to Claudia Maraston, Daniel Thomas, David Elbaz, Fabio Vito and Sirio Belli for helpful discussions. This study was supported by the LabEx UnivEarthS funding. R.Gobat acknowledges funding from project ANID Fondecyt 1231661. This research is based on observations made with the NASA/ESA James Webb Space Telescope obtained from the Space Telescope Science Institute, which is operated by the Association of Universities for Research in Astronomy, Inc., under NASA contract NAS 5–26555. These observations are associated programs ID\#01207, PI: G. Rieke; ID\#06541, PI: E. Egami; ID\#06549, PI: J. Pierel; ID\#03547, PI: H. Umehata. This paper makes use of the following ALMA data: ADS/JAO.ALMA\#2019.1.00227.S, PI: K. Whitaker; ADS/JAO.ALMA\#2023.1.00609.S, PI: R. Gobat; ADS/JAO.ALMA\#2023.1.01016.S, PI: Z. Ji; ADS/JAO.ALMA\#2024.1.00099.S, PI: R. Gobat. ALMA is a partnership of ESO (representing its member states), NSF (USA) and NINS (Japan), together with NRC (Canada), MOST and ASIAA (Taiwan), and KASI (Republic of Korea), in cooperation with the Republic of Chile. The Joint ALMA Observatory is operated by ESO, AUI/NRAO and NAOJ.

\bibliographystyle{aa}
\bibliography{references.bib}

\nocite{Williams21}
\nocite{Suess17}
\nocite{Belli21}
\nocite{Rudnick17}
\nocite{Hayashi18}
\nocite{Spilker18}
\nocite{Zanella23}
\nocite{Sargent15}
\nocite{Caliendo21}
\nocite{Scholtz24}
\nocite{zavala19}
\nocite{Donevski23}
\nocite{Suess25}
\nocite{Parente26}

\appendix

% \section{QG2 MIRI/F2100W residuals}
% We here include QG2's MIRI/F2100W residuals, after subtracting the central stellar component. We overlay the [CII] contours of this source. We find good spatial agreement between MIRI residuals and [CII] emission, suggesting that [CII] is tracing diffuse warm dust in QG2. This is qualitatively consistent with the idea that the mechanism that heats QG2's dust is also boosting its [CII] line.

% \begin{figure}[]
%  \centering
%     \includegraphics[width=\columnwidth]{QG2_residuals_F2100W.pdf}%
%     % \caption{2 Figures side by side}%%
%     \caption{Background image: JWST/MIRI F2100W residuals for QG2. [CII] contours are drawn as in Fig.~\ref{fig:all}. The red cross marks the position of QG2's stellar core. The cutout has a size of 3".}
%     \label{fig:QG2_MIRI}
% \end{figure}

\section{QG1-Sat [CII] spectrum}
We here include the spectrum of QG1-Sat, a [CII] source found 2'' offset from QG1.
\begin{figure}[]
 \centering
    \includegraphics[width=\columnwidth]{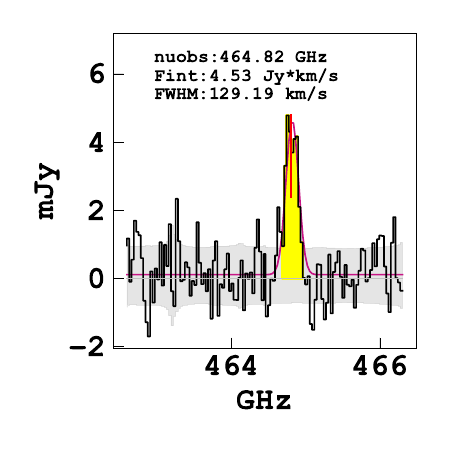}%
    % \caption{2 Figures side by side}%%
    \caption{[CII] spectrum extracted on QG1-Sat. The red vertical line marks the expected observed-frame frequency of [CII] based on the spectroscopic redshift from the CO emission line reported in \cite{Umehata2025}. The magenta solid line marks the best fit Gaussian model from MPFIT.}
    \label{fig:QG1_offset}
\end{figure}

\section{MICHI2 plots}
\label{sec:michi2_plots}
In the following Figures, for each source, we show templates included in the 1$\sigma$ confidence level in our MICHI2 fits (left panels), and the $\chi^2$ distribution of the varying parameters in the corresponding fits (right panels). 

\begin{figure*}
\centering
\begin{subfigure}{.55\textwidth}
  \centering
  \includegraphics[width=.8\textwidth]{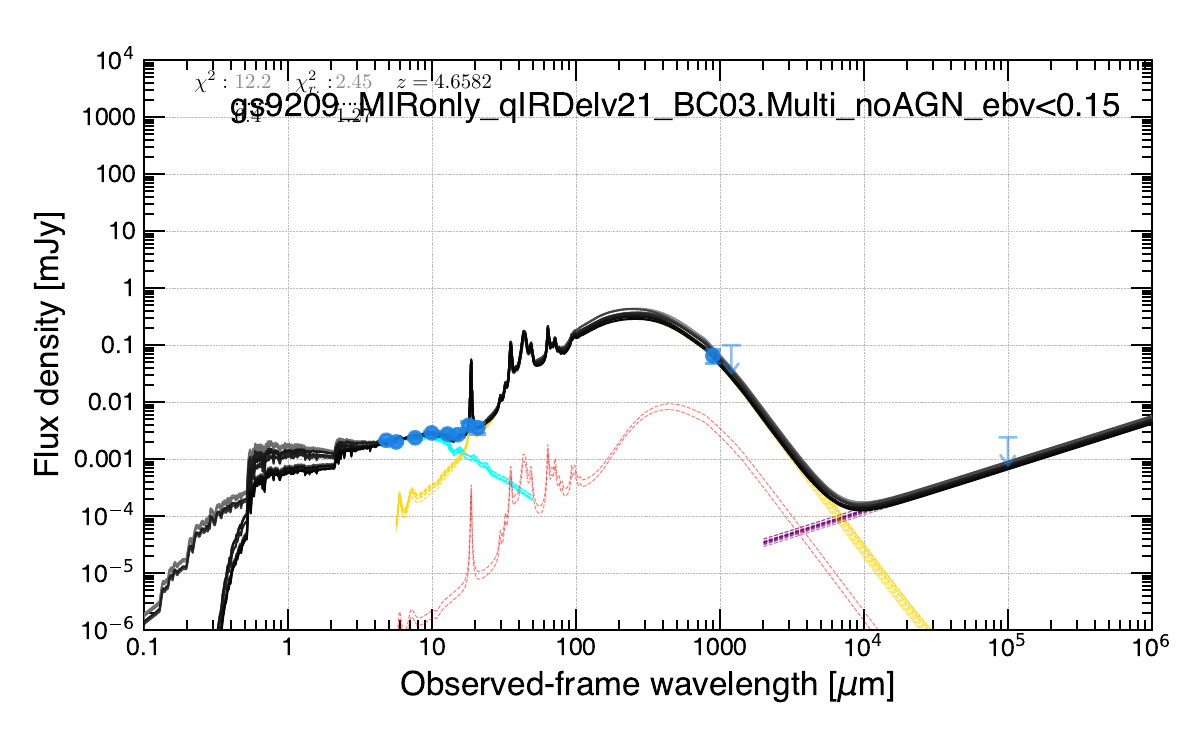}
  \label{fig:sub1}
\end{subfigure}%
\begin{subfigure}{.45\textwidth}
  \centering
  \includegraphics[width=.8\textwidth]{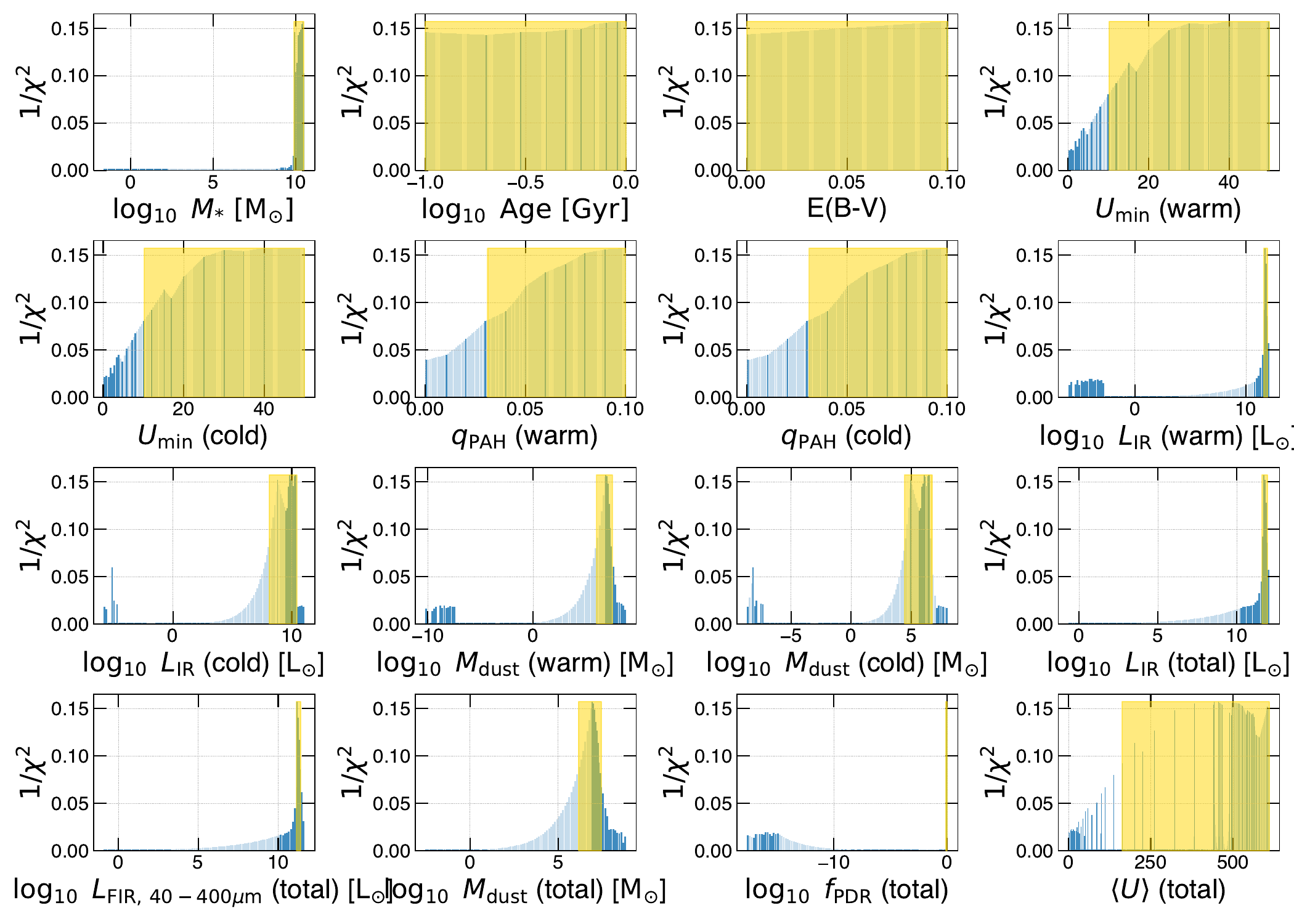}
  \label{fig:sub2}
\end{subfigure}
\caption{Left: GS-9209 without an AGN torus component with E(B-V)$<$0.15 mag. Gray lines show solutions within 1-sigma of the best fit (i.e. minimum-$\chi^2$ solution), marked as a solid black line. The cyan solid line marks the stellar component. The yellow, red and purple lines mark the warm dust, cold dust and synchrotron radio components, respectively. Right: $\chi^2$ distribution for the free parameters of the fit. The yellow region marks values included in the 1$\sigma$ confidence range.}
\label{fig:gs_sed_noagn}
\end{figure*}

\begin{figure*}
\centering
\begin{subfigure}{.55\textwidth}
  \centering
  \includegraphics[width=.8\textwidth]{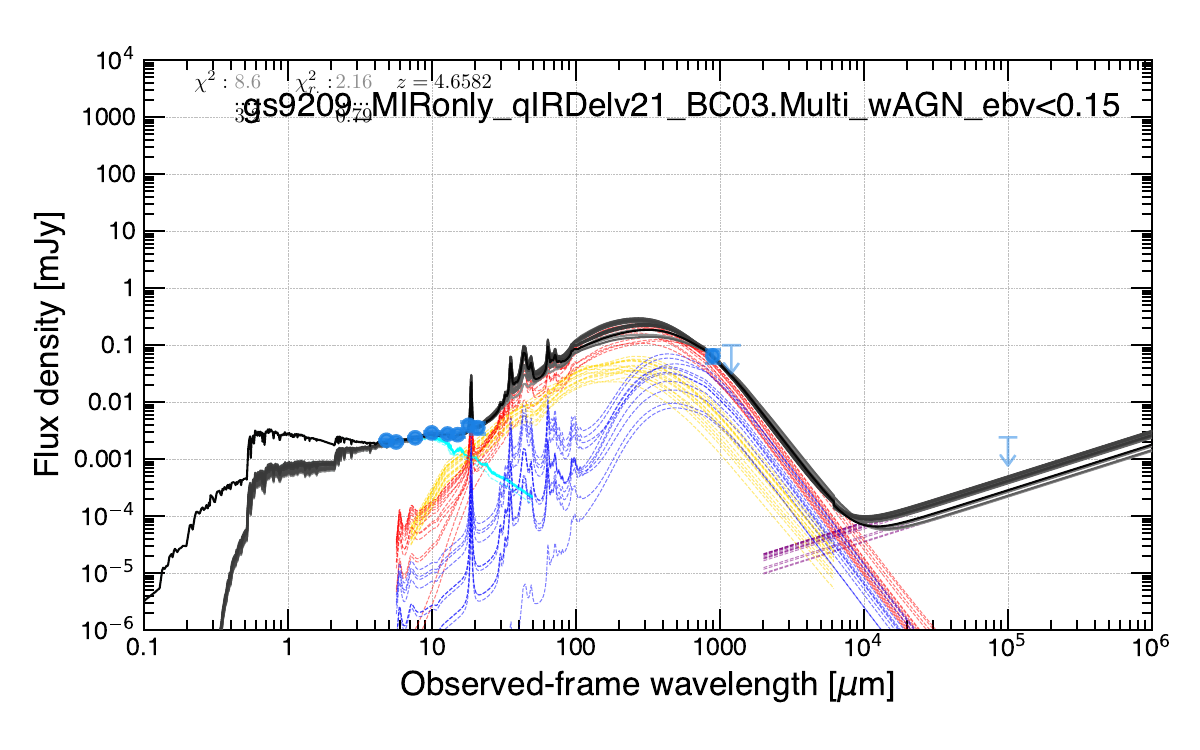}
  \label{fig:sub1}
\end{subfigure}%
\begin{subfigure}{.45\textwidth}
  \centering
  \includegraphics[width=.8\textwidth]{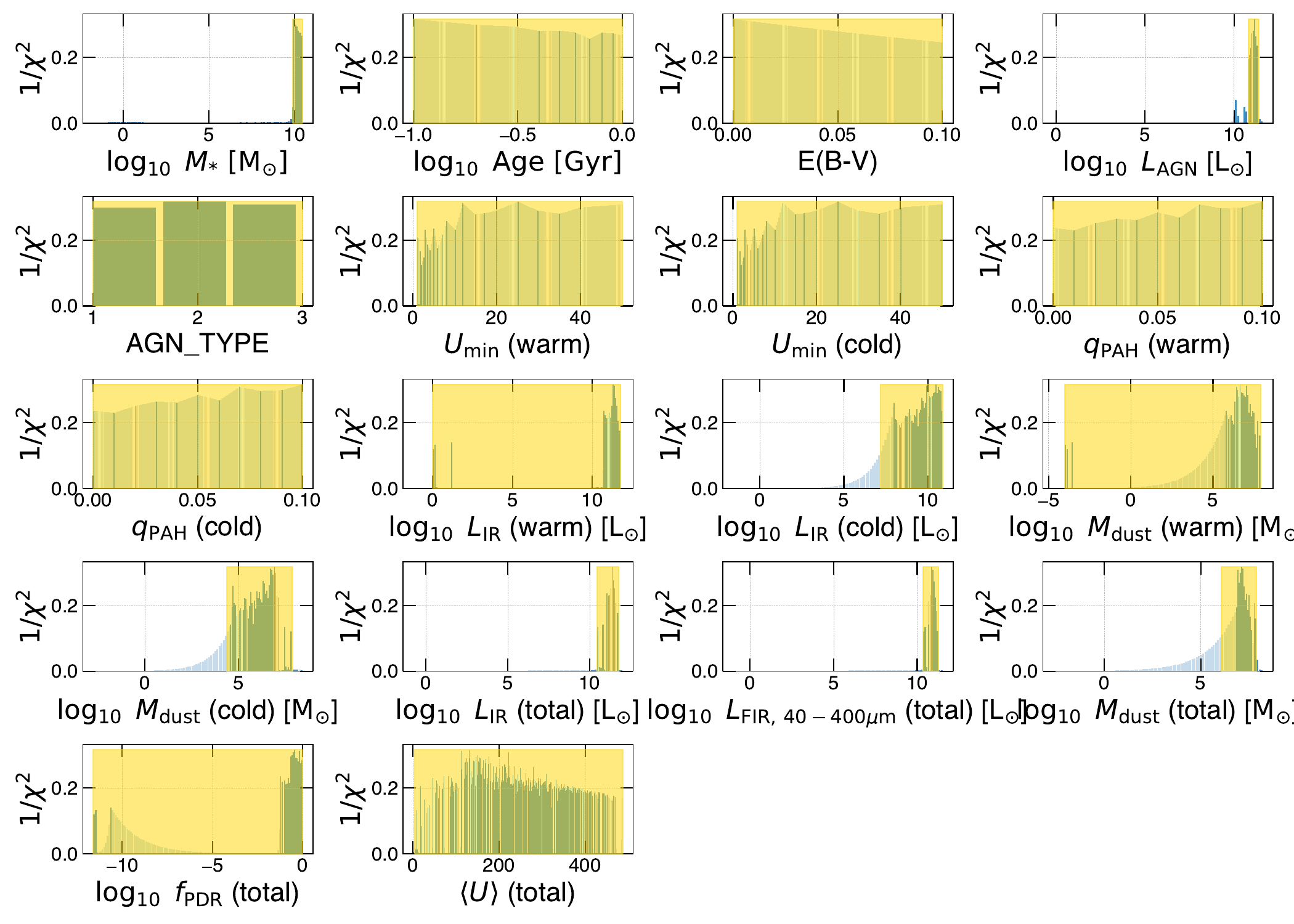}
  \label{fig:sub2}
\end{subfigure}
\caption{Same as Fig.~\ref{fig:gs_sed_noagn} for GS-9209 but including an AGN torus component. In this case, the yellow, red and blue lines mark the AGN, warm dust and cold dust components, respectively. The cyan and purple solid lines mark the stellar and radio templates, respectively.}
\label{fig:gs_sed_agn}
\end{figure*}

\begin{figure*}
\centering
\begin{subfigure}{.55\textwidth}
  \centering
  \includegraphics[width=.8\textwidth]{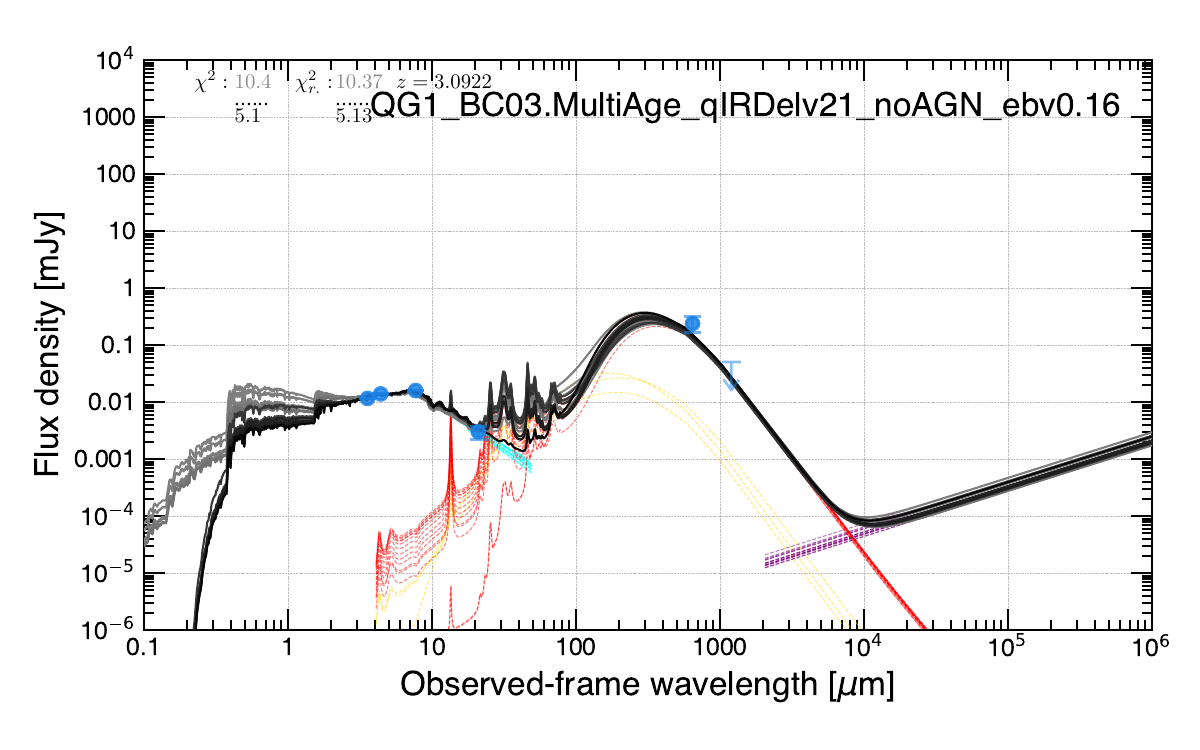}
  \label{fig:sub1}
\end{subfigure}%
\begin{subfigure}{.45\textwidth}
  \centering
  \includegraphics[width=.8\textwidth]{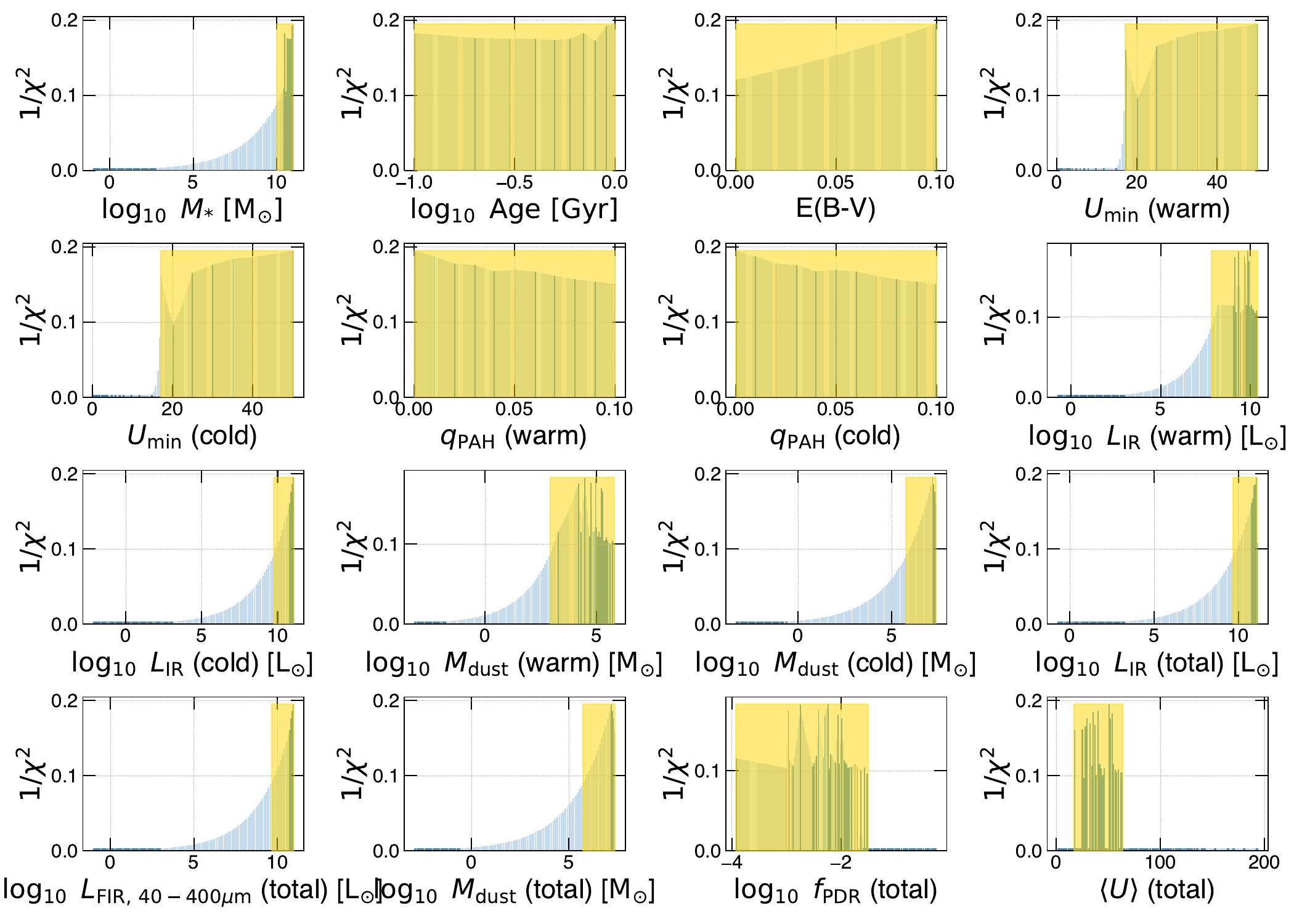}
  \label{fig:sub2}
\end{subfigure}
\caption{Same as Fig.~\ref{fig:gs_sed_noagn} but for QG1 without an AGN template.}
\label{fig:qg1sed_noAGN}
\end{figure*}

\begin{figure*}
\centering
\begin{subfigure}{.55\textwidth}
  \centering
  \includegraphics[width=.8\textwidth]{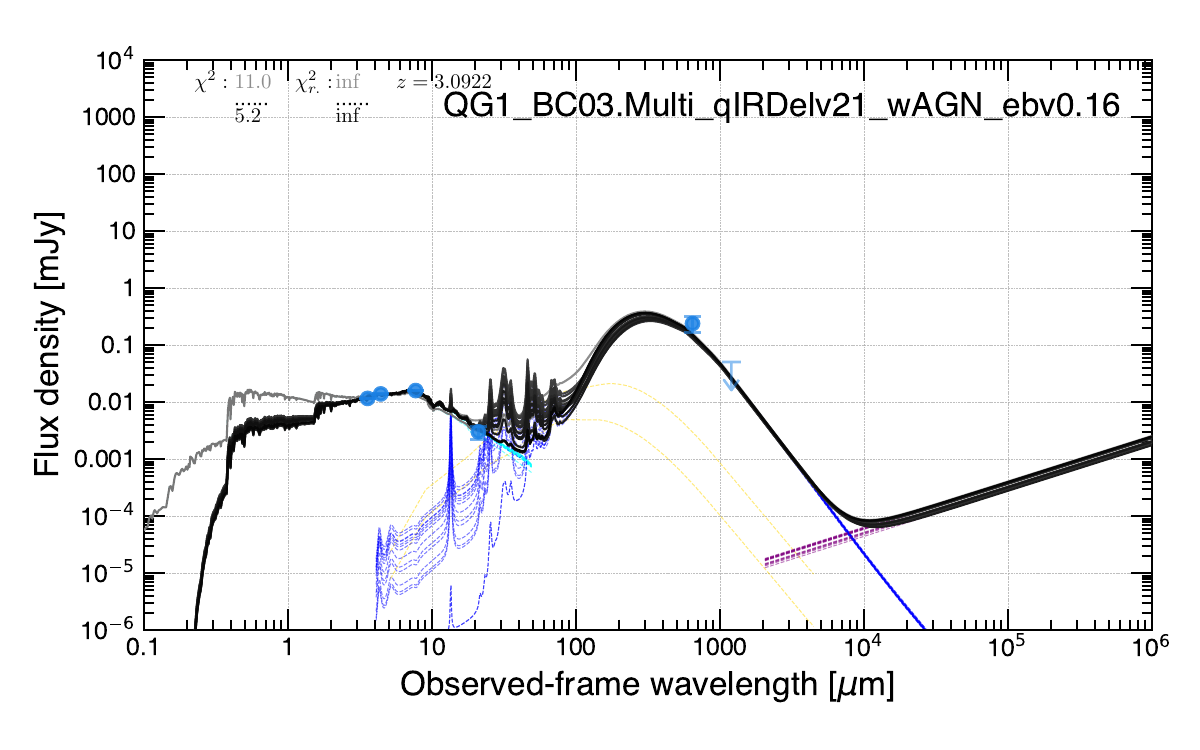}
  \label{fig:sub1}
\end{subfigure}%
\begin{subfigure}{.45\textwidth}
  \centering
  \includegraphics[width=.8\textwidth]{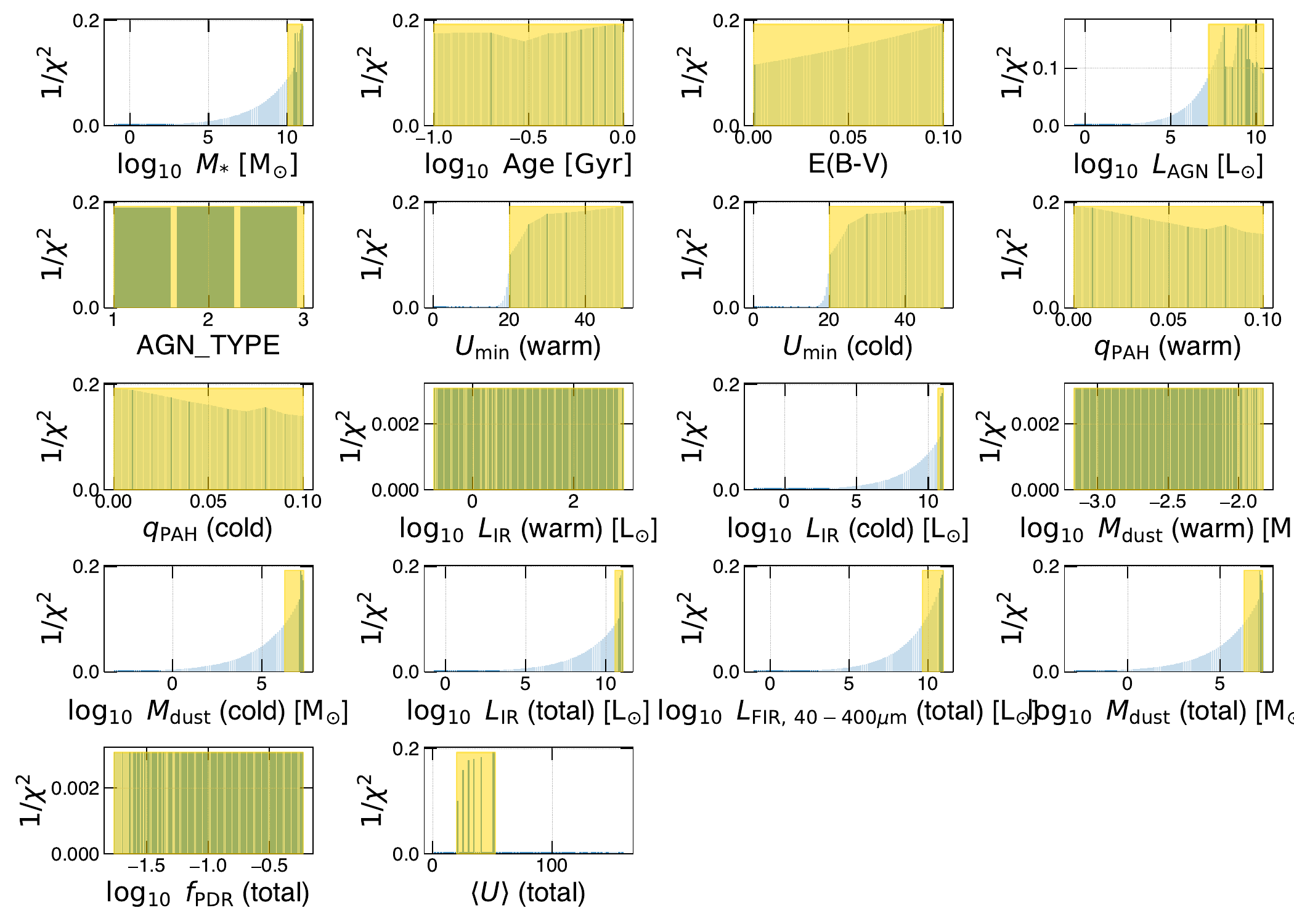}
  \label{fig:sub2}
\end{subfigure}
\caption{Same as Fig.~\ref{fig:qg1sed_noAGN} but including an AGN template.}
\label{fig:qg1sed_wAGN}
\end{figure*}

\begin{figure*}
\centering
\begin{subfigure}{.55\textwidth}
  \centering
  \includegraphics[width=.8\textwidth]{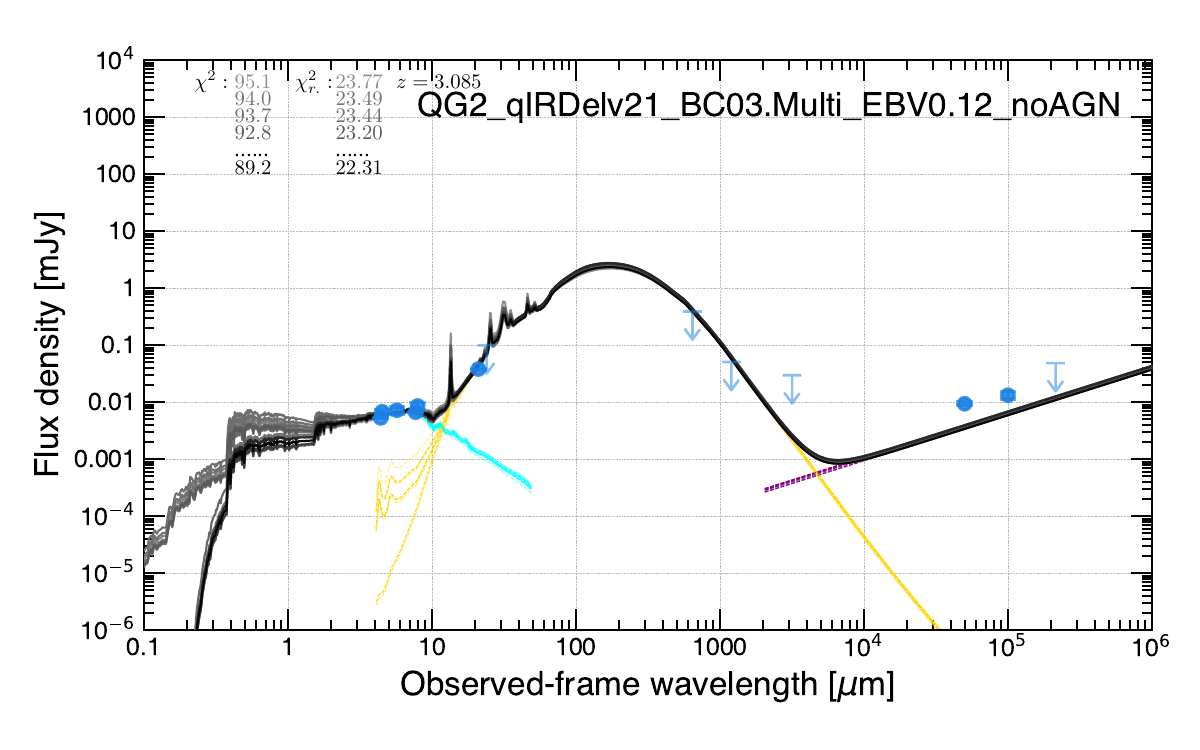}
  \label{fig:sub1}
\end{subfigure}%
\begin{subfigure}{.45\textwidth}
  \centering
  \includegraphics[width=.8\textwidth]{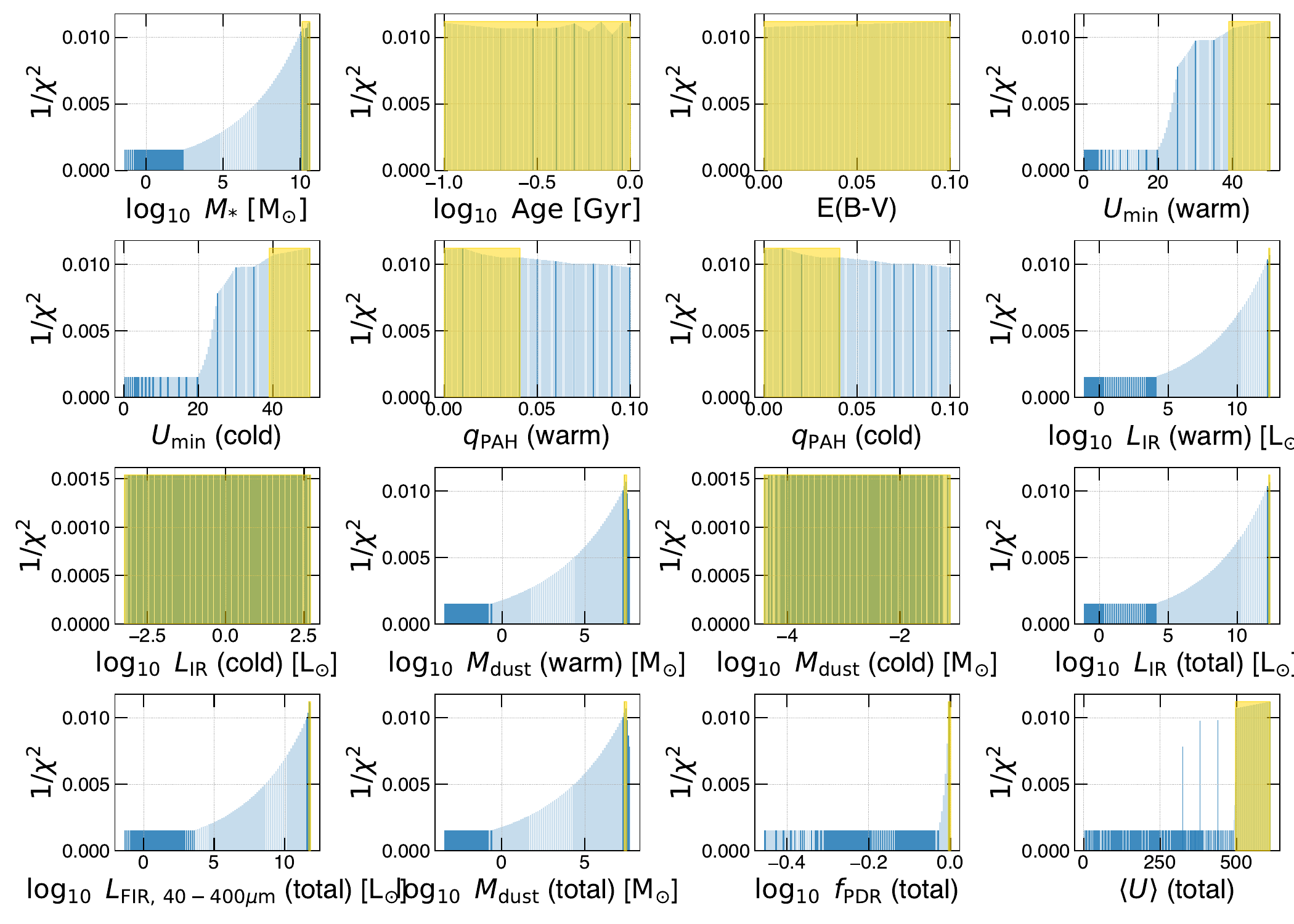}
  \label{fig:sub2}
\end{subfigure}
\caption{Same as Fig.~\ref{fig:gs_sed_noagn} but for QG2 without an AGN template and with E(B-V)$<$0.12 mag.}
\label{fig:qg2sed_noAGN}
\end{figure*}

\begin{figure*}
\centering
\begin{subfigure}{.55\textwidth}
  \centering
  \includegraphics[width=.8\textwidth]{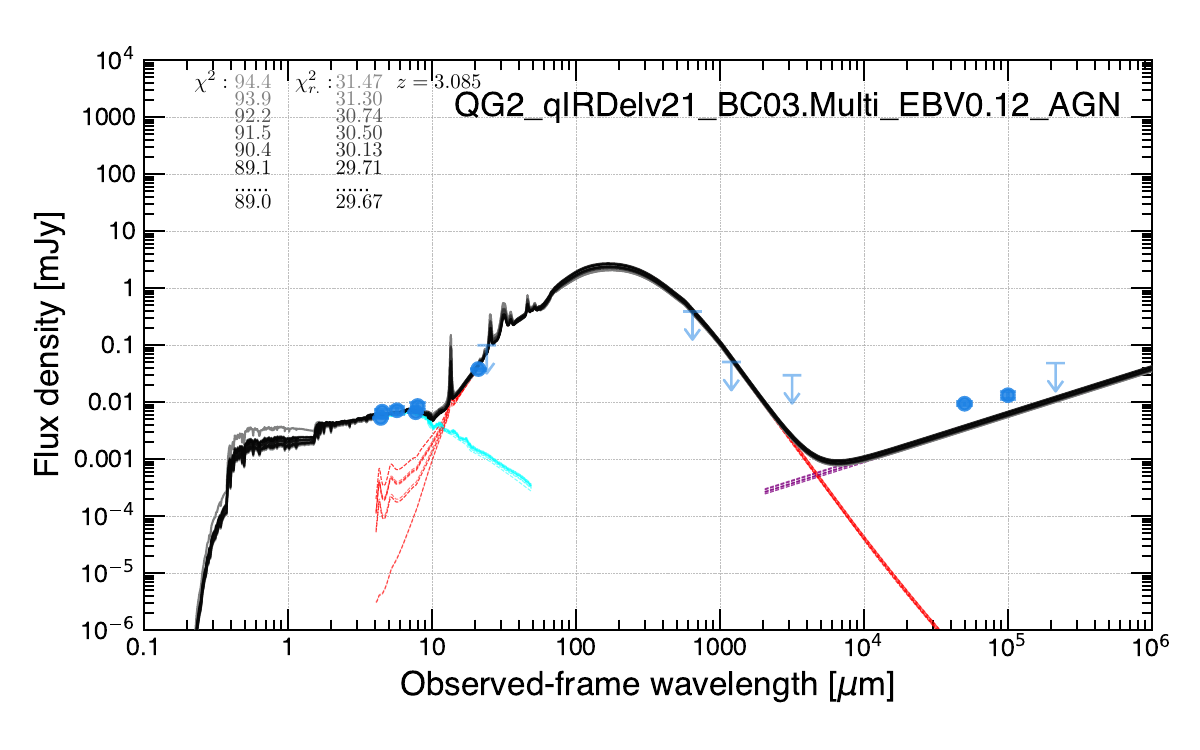}
  \label{fig:sub1}
\end{subfigure}%
\begin{subfigure}{.45\textwidth}
  \centering
  \includegraphics[width=.8\textwidth]{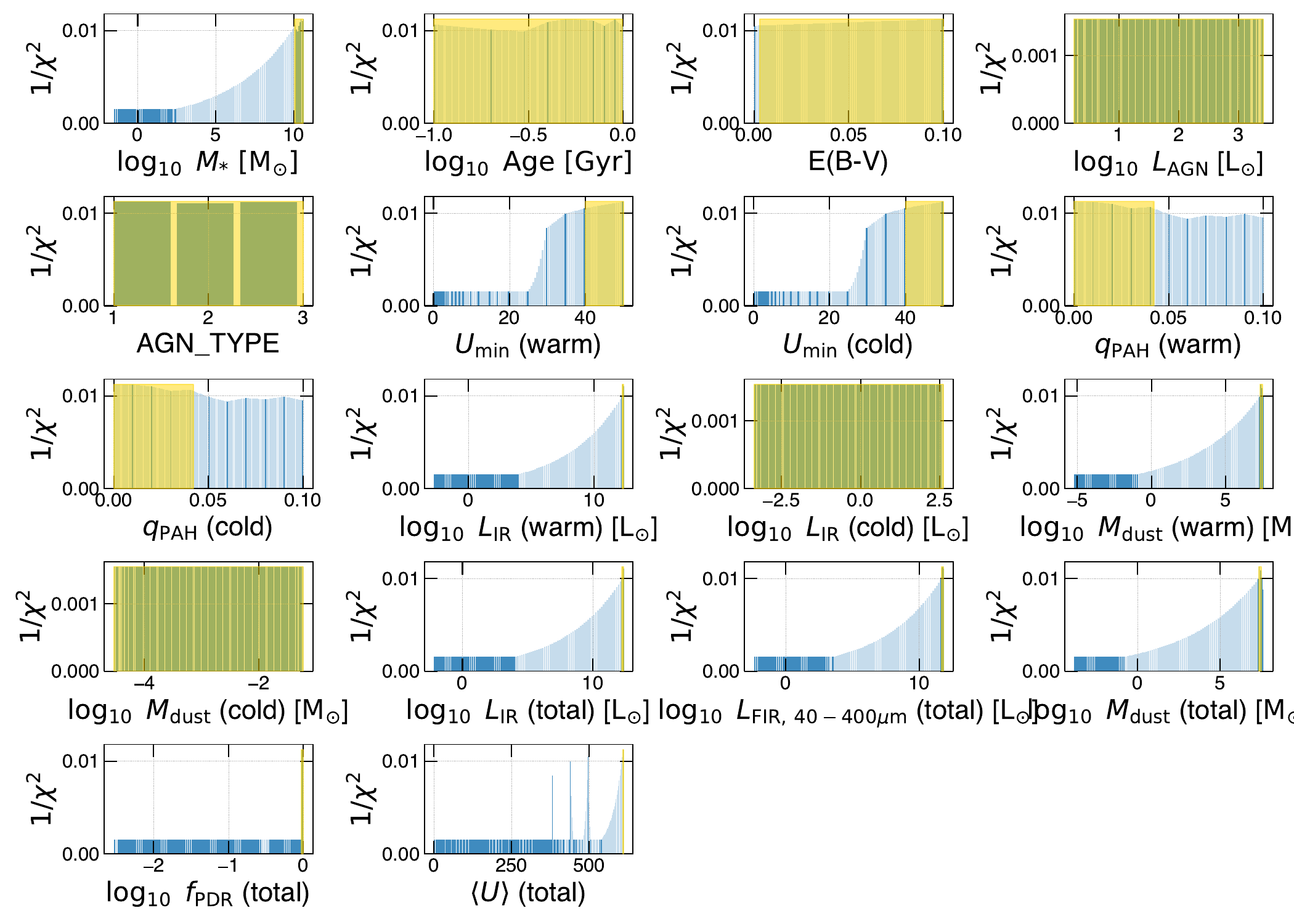}
  \label{fig:sub2}
\end{subfigure}
\caption{Same as Fig.~\ref{fig:qg2sed_noAGN} but including an AGN template.}
\label{fig:qg2sed_wAGN}
\end{figure*}

\begin{figure*}
\centering
\begin{subfigure}{.55\textwidth}
  \centering
  \includegraphics[width=.8\textwidth]{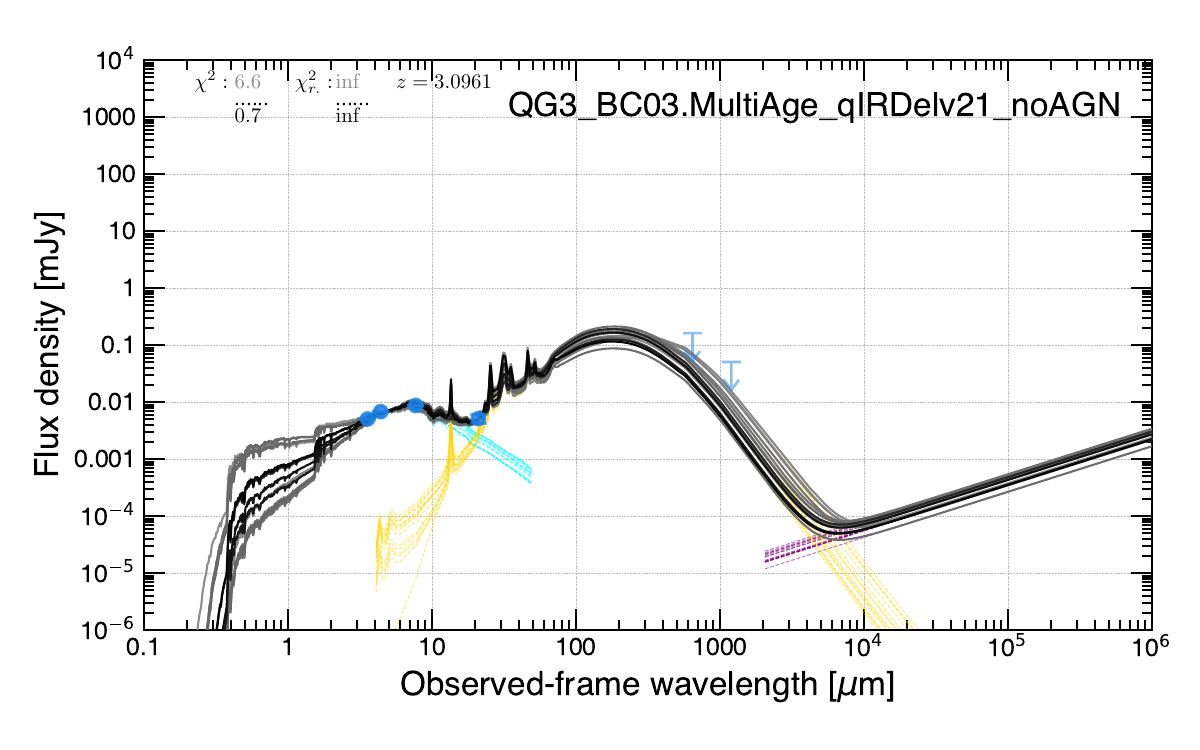}
  \label{fig:sub1}
\end{subfigure}%
\begin{subfigure}{.45\textwidth}
  \centering
  \includegraphics[width=.8\textwidth]{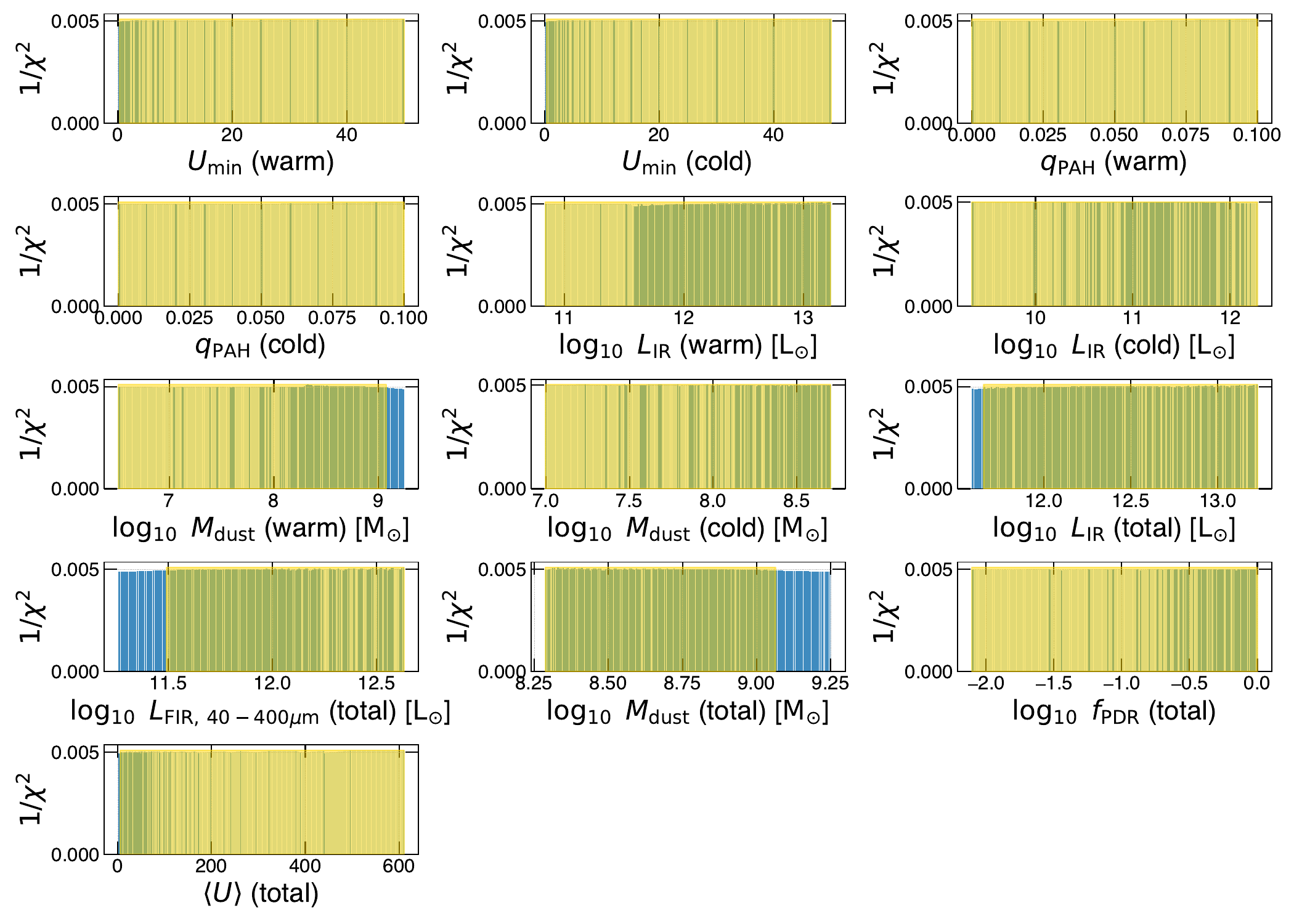}
  \label{fig:sub2}
\end{subfigure}
\caption{Same as Fig.~\ref{fig:gs_sed_noagn} but for QG3 without an AGN template.}
\label{fig:qg3sed_noAGN}
\end{figure*}

\begin{figure*}
\centering
\begin{subfigure}{.55\textwidth}
  \centering
  \includegraphics[width=.8\textwidth]{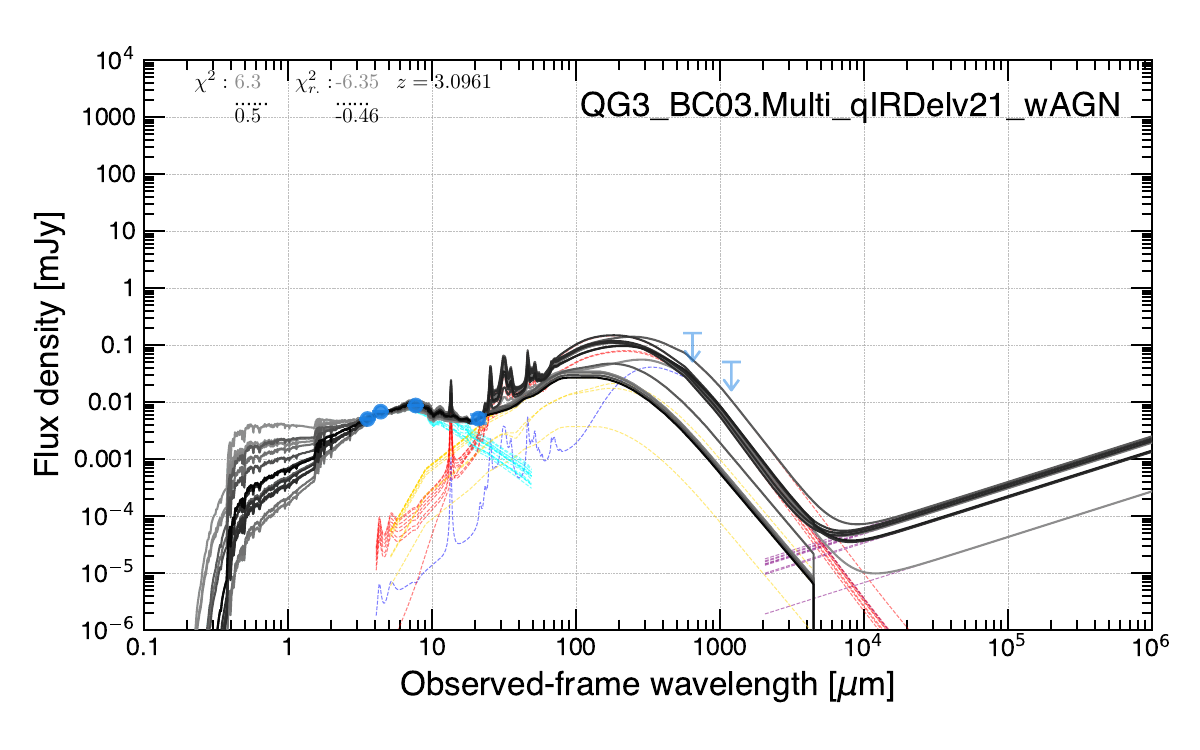}
  \label{fig:sub1}
\end{subfigure}%
\begin{subfigure}{.45\textwidth}
  \centering
  \includegraphics[width=.8\textwidth]{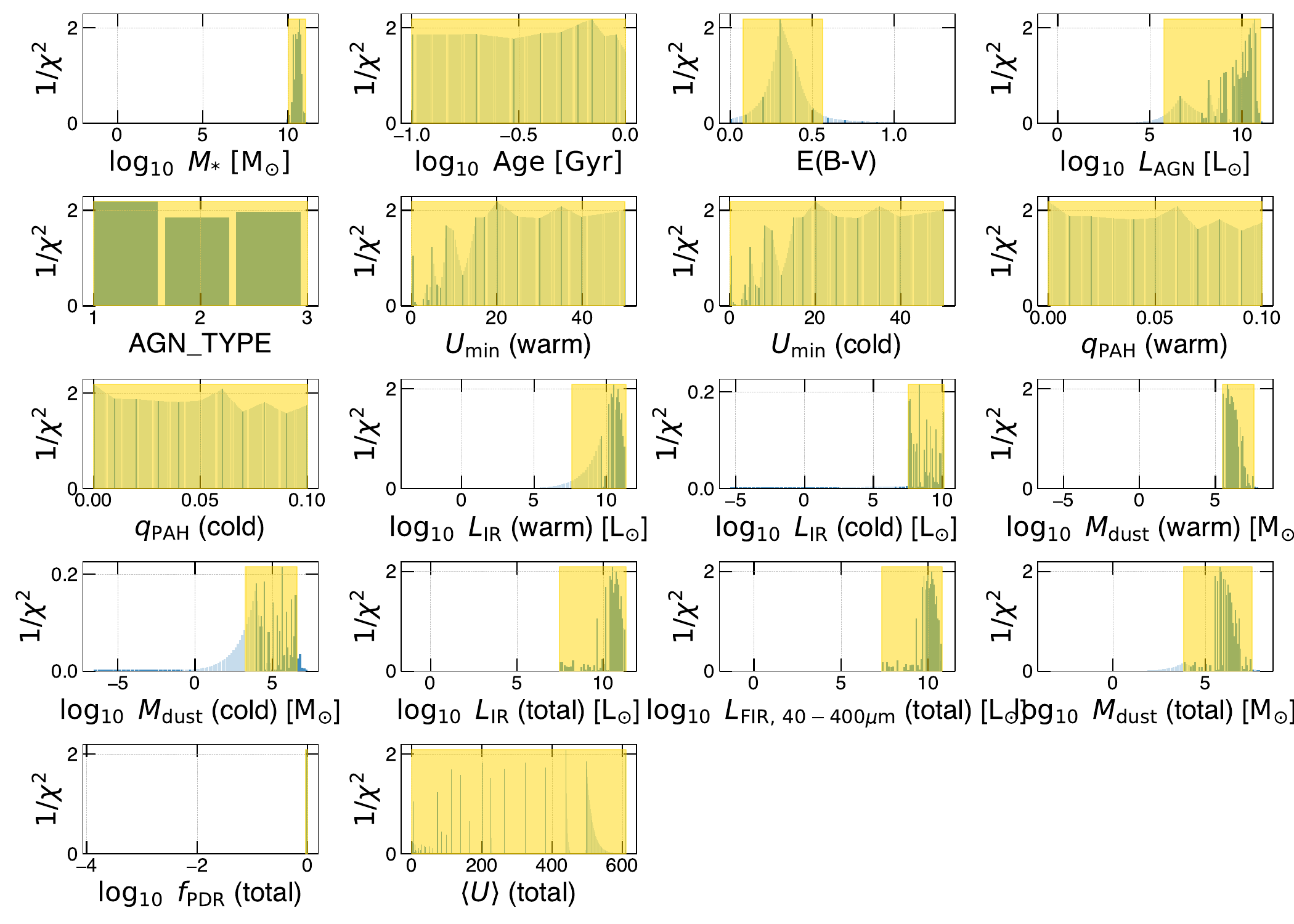}
  \label{fig:sub2}
\end{subfigure}
\caption{Same as Fig.~\ref{fig:qg3sed_noAGN} but including an AGN template.}
\label{fig:qg3sed_wAGN}
\end{figure*}

\begin{figure*}
\centering
\begin{subfigure}{.55\textwidth}
  \centering
  \includegraphics[width=.8\textwidth]{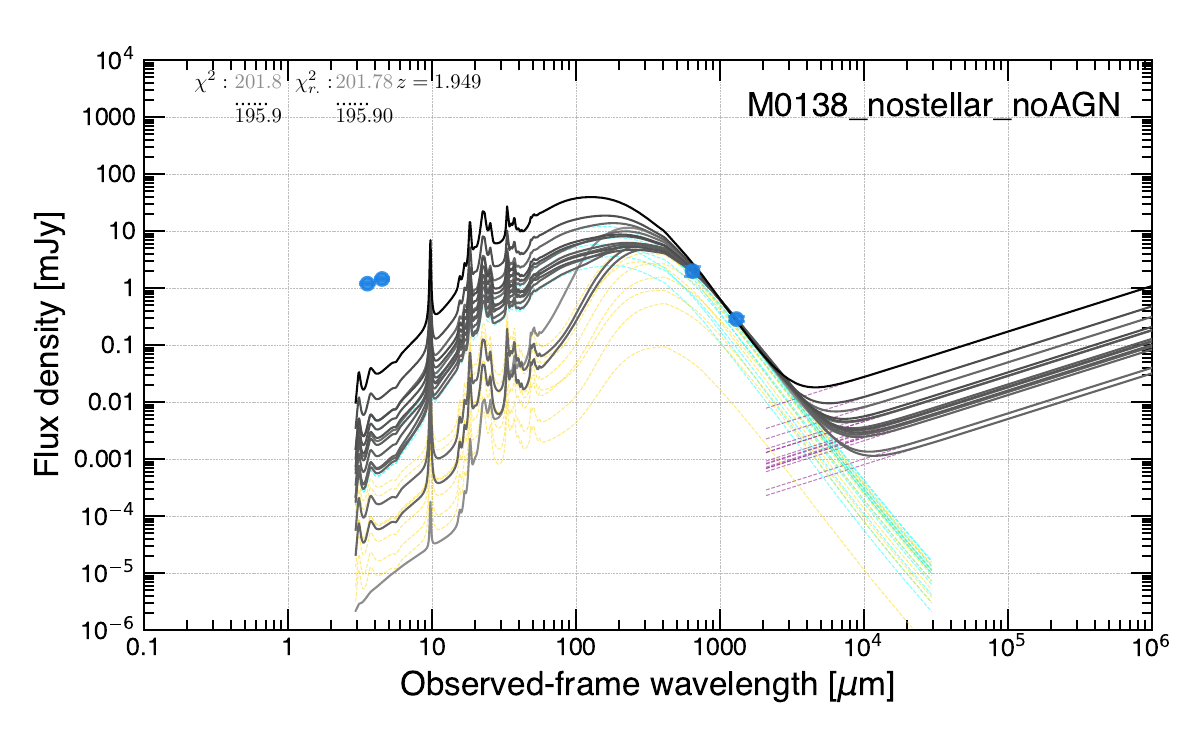}
  \label{fig:sub1}
\end{subfigure}%
\begin{subfigure}{.45\textwidth}
  \centering
  \includegraphics[width=.8\textwidth]{fit_5.chisq_SED_M0138_nostellar_noAGN.pdf}
  \label{fig:sub2}
\end{subfigure}
\caption{Same as Fig.~\ref{fig:gs_sed_noagn} but for M0138 without an AGN template. No stellar component was fitted.}
\label{fig:m0138sed}
\end{figure*}

\end{document}